\documentclass[usenatbib]{mn2e}
\usepackage{color}
\usepackage{natbib}
\usepackage{paralist}
\usepackage{graphicx}
\usepackage{epstopdf}
\usepackage{epsfig}
\usepackage{amsmath}
\usepackage{tablefootnote}

\usepackage{amssymb}
\usepackage{lipsum}
\usepackage{float}
\usepackage{array}
\usepackage{multirow}
\usepackage{setspace}

\title[Mapping the aliphatic hydrocarbon content of ISD]{A method for mapping the aliphatic hydrocarbon content of interstellar dust towards the Galactic Centre}
\author[B. G\"{u}nay, M. G. Burton, M. Af\c{s}ar, T. W. Schmidt]
{B. G\"{u}nay $^{1,3,4}$\thanks{E-mail: burcu.gunay@ege.edu.tr (BG);
mgb@arm.ac.uk (MGB)},
M. G. Burton$^{2,3}$, M. Af\c{s}ar$^{1}$, T. W. Schmidt$^{4}$ \\
\\
$^{1}$Department of Astronomy and Space Sciences,
                 Ege University, 35100 Bornova, \.{I}zmir, Turkey\\
$^{2}$Armagh Observatory and Planetarium,
                 College Hill, Armagh, BT61 9DG, Northern Ireland, UK\\
$^{3}$School of Physics,
                 UNSW Sydney, NSW 2052, Australia\\
$^{4}$ARC Centre of Excellence in Exciton Science, School of Chemistry,
                 UNSW Sydney, NSW 2052, Australia\\
}

\begin{document}

\maketitle
\label{firstpage}
\begin{abstract}
In the interstellar medium, the cosmic elemental carbon abundance includes the total carbon in both gas and solid phases. The aim of the study was to trial a new method for measuring the amount and distribution of aliphatic carbon within interstellar dust over wide fields of view of our Galaxy. This method is based on measurement of the 3.4\,$\mu$m absorption feature from aliphatic carbonaceous matter. This can readily be achieved for single sources using IR spectrometers. However, making such measurements over wide fields requires an imaging IR camera, equipped with narrow-band filters that are able to sample the spectrum. While this cannot produce as good a determination of the spectra, the technique can be applied to potentially tens to hundreds of sources simultaneously, over the field of view of the camera. We examined this method for a field in the centre of the Galaxy, and produced a map showing the variation of 3.4\,$\mu$m optical depth across it.
\end{abstract}


\section{Introduction}\label{Section1}

	The element carbon is key for probing the life cycle of the gas in the interstellar medium (ISM), as its major component, hydrogen is largely invisible, when in molecular form. When a part of the gas phase, carbon can be observed using high frequency radio telescopes, operating in the millimetre and terahertz bands. However, there is a missing ingredient in this case, carbon in the solid phase, locked in interstellar dust. To be able to measure the distribution of carbon in solid phase, a new observation method is required to map the carbonaceous interstellar dust across the Galaxy, as is laboratory research to quantify the amount of carbon incorporated in interstellar dust.

	Carbon is the 4th most abundant element in the ISM. It is chemically versatile to form molecules. Carbon is able to bond through three different orbital hybridisations; sp$^{3}$, sp$^{2}$ and sp and can form aliphatics (alkane, sp$^3$), olefinics (alkene, sp$^2$), aromatics (sp$^2$) and acetylenics (alkynes, sp). Therefore, there is a wide diversity of carbon compounds and rich carbon chemistry in the ISM.

	There are prominent spectral features of carbonaceous dust in spectra of the ISM in the infrared region. These absorption features are at 3.28\,$\mu$m, 3.4\,$\mu$m, 5.87\,$\mu$m, 6.2\,$\mu$m, 6.85\,$\mu$m and 7.25\,$\mu$m \citep{Dartois2004}. The 3.4\,$\mu$m absorption feature is of particular interest since it is more prominent and prevalent towards background IR radiation sources.

	The 3.4\,$\mu$m absorption feature has been extensively observed through several sightlines toward the Galactic Centre \citep{Soifer1976, Willner1979, Wickramasinghe1980, Butchart1986, McFadzean1989, Tielens1996, Adamson1990, Sandford1991, Pendleton1994, Chiar2000, Chiar2002, Chiar2013, Moultaka2004}, Local ISM \citep{Butchart1986, Adamson1990, Sandford1991, Pendleton1994, Whittet1997}, planetary nebula \citep{Lequeux1990, Chiar1998} and the ISM of other galaxies \citep{Imanishi2000, Mason2004, Dartois2004, Geballe2009}. The 3.4\,$\mu$m feature has also been detected in the spectra of Solar System materials such as meteorites (e.g. \citealt{Ehrenfreund1991}), Interplanetary Dust Particles (IDPs) (e.g. \citealt{Matrajt2005}) and cometary grains (e.g. \citealt{MunozCaro2008}).

	The observed 3.4\,$\mu$m absorption is attributed to the aliphatic C-H stretch in carbonaceous dust. Therefore, the optical depth of the 3.4\,$\mu$m absorption feature is related to the number of aliphatic carbon C-H bonds along the line of sight. It is possible to estimate the column density ($N$, cm$^{-2}$) of carbon locked up in aliphatic hydrocarbon material of interstellar dust grains from quantitative analysis of the 3.4\,$\mu$m interstellar absorption feature. 
	
	One can measure the optical depth ($\tau$) of the 3.4\,$\mu$m (2940 cm$^{-1}$) absorption and determine the number of C-H bonds needed to produce this absorption by using the laboratory measurements of the integrated absorption coefficient ($A$, cm molecule$^{-1}$) and line width ($\Delta \bar{\nu}$, cm$^{-1}$) (which, for the low resolution spectra we use, becomes the filter bandwidth).

\begin{equation}\label{eq:1}
N =\frac{\tau \Delta \bar{\nu}} {A}
\end{equation}

This is the second of two papers concerning the measurement of the aliphatic hydrocarbon component in interstellar dust.  The first paper (\citealt{Gunay2018}, hereafter Paper 1) undertook laboratory measurements in order to provide a revised value for the absorption coefficient of aliphatic hydrocarbons. Interstellar dust analogues (ISDAs) were produced under simulated circumstellar/interstellar conditions and providing that their spectra are consistent with the interstellar spectra, they were used for the measurements. We treat the aliphatic absorption as a single feature and have determined the integrated absorption coefficient for $-$CH$_{x}$ groups consisting of a mixture of aliphatic $-$CH$_{2}$ and $-$CH$_{3}$ groups. The resultant integrated absorption coefficients obtained by employing amorphous aliphatic hydrocarbon containing ISDAs were found to be less than half those mesaured using small aliphatic hydrocarbon molecules \citep{Sandford1991, Dartois2007}.
		
Using the absorption coefficient of 4.69 $\times$ 10$^{-18}$\,cm group$^{-1}$ and reappraising the spectroscopic observations of \cite{Pendleton1994} for Galactic Centre source GCIRS 6E, we obtained an aliphatic carbon column density of $4.87\times10^{18}$\,cm$^{-2}$, which is a factor of five higher than previously reported by \cite{Pendleton1994} for the same sightline. Similarly, employing $\tau_{3.4\,\mu m}$ value of GCIRS 6E from \citet{Chiar2002} (hereafter [C02]) we found the aliphatic carbon column density of $5.86\times10^{18}$\,cm$^{-2}$.
We calculated the aliphatic carbon abundances in ppm\footnote{ppm: parts per million} towards a range of Galactic Centre sources (Table 3 in Paper 1) using $\tau_{3.4\,\mu m}$ values obtained from [C02]. Assuming A$_{V}$$\sim$30 mag \citep{Becklin1978}, normalised aliphatic hydrocarbon abundances (C/H) (ppm) were calculated based on gas-to-extinction ratio $N(H) = 2.04 \times 10^{21}$\,cm$^{-2}$ mag$^{-1}$ \citep{Zhu2017}. The minimum and maximum normalised aliphatic carbon abundances were obtained as 55$-$135\,ppm. The average value for the Galactic Centre was found to be 93\,ppm based on the sources taken into account. 
			
The ISM carbon abundance derived from Solar atmosphere \citep{Grevesse1998,Asplund2005,Asplund2009}, and meteoritic/protosolar abundances \citep{Lodders2003} is up to 270\,ppm carbon. The ISM carbon abundance has been studied by using B type stars and found to be $209\pm15$\,ppm \citep{Sofia2001, Przybilla2008}. \cite{Sofia2001} found the total ISM carbon abundance around $358\pm82$\,ppm from the atmosphere of young F, G type stars. Therefore the average value of 93\,ppm aliphatic hydrocarbon corresponds to at least a quarter of the available carbon in the ISM \citep{Sofia2001}. This result indicates that a significant part of the ISM carbon is incorporated in aliphatic content of ISM dust.

On the other hand, there are discrepancies in the aliphatic hydrocarbon abundances arising from $\tau_{3.4\,\mu m}$ values reported in the literature. For example, [C02] reported $\tau_{3.4\,\mu m}$ = 0.147 towards GCIRS 7, which yields only 55\,ppm aliphatic hydrocarbon, where as \cite{Moultaka2004} (hereafter [M04]) reported $\tau_{3.4\,\mu m}$ = 0.41 towards GCIRS 7, which yields 150\,ppm aliphatic hydrocarbon. There are also uncertainties in $\tau_{3.4\,\mu m}$ values arising from analysis methods of $\tau$ in the literature. [M04] highlighted that $\tau_{3.4\,\mu m}$ values differ for each line of sight based on the continuum fit applied. They reported that $\tau_{3.4\,\mu m}$ = 0.49 towards GCIRS 16C (the maximum $\tau_{3.4\,\mu m}$ reported in the literature), which yields 179\,ppm aliphatic hydrocarbon. They also obtained the lower limit of $\tau_{3.4\,\mu m}$ = 0.14 for the same line of sight using a linear continuum, which yields only 51\,ppm aliphatic hydrocarbon. As a consequence, $\tau_{3.4\,\mu m}$ values can depend on the measurement and analysis method. To obtain relative abundances of aliphatic hydrocarbon in ISM dust, we need consistent $\tau_{3.4\,\mu m}$ values obtained using the same methodology through the all sources in the field of interest.

To map the aliphatic hydrocarbon abundances incorporated in ISM dust, we need to measure $\tau_{3.4\,\mu m}$ for as many sources as possible in a field of interest. The measurement of the optical depth can readily be achieved for single sources using IR spectrometers. However, making such measurements over wide fields with many sources requires long observing times.
		
Instead, spectrophotometry can be employed: an imaging IR camera, equipped with narrow-band filters to sample the spectrum. The photometric measurements obtained with narrow-band filters then can be used to build low-resolution spectra. Although this technique would not produce as good a determination of the absorption feature as spectra for a single source, it may be applied to potentially tens to hundreds of sources simultaneously, over the field of view of the camera. Therefore this technique enables us to measure the optical depth of the 3.4\,$\mu$m absorption feature for large fields with a low cost in observing time. Optical depth of the 3.4\,$\mu$m absorption feature will be used to find the column density of the aliphatic carbon, by using the integrated absorption coefficient obtained by laboratory measurements.

This paper is organised as follows. The observations are described and the strategies are presented in Section \ref{Section2}. The details of the data reduction and data analysis are summarised in Section \ref{Section3} and \ref{Section4}. Results are reported in Section \ref{Section5}. Discussion and conclusions are presented in Section \ref{Section6}.

\section{Observations}\label{Section2}
Observations were carried out with the UIST\footnote{https://www.ukirt.hawaii.edu/instruments/uist/uist.html} instrument on the 3.8\,m United Kingdom Infrared Telescope (UKIRT). We summarize the features of the observed field and the observing log in Table \ref{tab:1}. 

Spectrophotometric measurements were obtained using narrow-band filters\footnote{http://www.ukirt.hawaii.edu/instruments/uist/imaging/filters.html} to build low-resolution spectra of stars towards the Galactic Centre. We aimed to determine how well the optical depth of the 3.4$\,\mu$m absorption feature could be measured by taking images using a suite of narrow-band filters spread across the 3.4$\,\mu$m L$-$band: 3.05 ice MK, 3.29 PAH MK, 3.4 nbL, 3.5 mbL, 3.6 nbL$'$, 3.99 (cont). Their 50$\%$ cut-on and 50$\%$ cut-off wavelengths, bandwidths and peak transmissions are presented in Table \ref{tab:2}.

The observing strategy was to image the Galactic Centre with a 3$\times$3 jitter pattern with a 20 arcsec grid. We applied 1 minute integration per jitter position. The pixel scale was 0.12 arcsec.

We calculated the sensitivity thresholds for S/N = 5 using the UIST online calculator\footnote{http://www.ukirt.hawaii.edu/cgi-bin/ITC/itc.pl} for 9 minutes of total integration time per filter (Table \ref{tab:2}). We applied 2 s exposure $\times$ 30 co-adds for the PAH MK, 3.4 nbL, 3.5mbL, 3.6 nbL$'$, 3.99  (cont) filters and 4 s exposure $\times$ 15 co-adds for the Ice MK filter (a longer exposure time can be used for the Ice MK due to the lower background compared to the other filters). However, observations could only be carried out using 5 filters (no 3.99 $\,\mu$m data) due to inclement weather conditions for the latter.

\begin{table}
 \begin{center}
  \caption{Galactic Coordinates of the centre of the observed field and the field of view.}
  \label{tab:1}
   \centering
  \begin{tabular}{| c | c |  }
  \hline
Target & Galactic Center   \\
  \hline

Longitude ($l$) & 359.945°  \\
Latitude ($b$) & -00.045°  \\
 \hline
FoV&163 arcsec  \\
 \hline
 \multirow{1}{*}{Jitter}&9pt jitter  \\
 \multirow{1}{*}{Pattern}&offsets of  $20''$ \\
  \hline
Observing Date & September 2015 \\

 \hline
\end{tabular}
\end{center}
\end{table}

\begin{table*}
\footnotesize
 \begin{center}

\centering
\caption{Filter set used.}

  \label{tab:2}
  \begin{tabular}{| c | c | c | c | c | c | c | c |c |}
    \hline
    
 \multirow{3}{*}{Filter}	&50$\%$  & 50$\%$   & FWHM & Peak    & Integration   & Sensitivity \\
&  Cut-On & Cut-Off  &  (Bandwidth) & Transmission & Time & Thresholds \\
&   ($\mu$m)	&   ($\mu$m) &  ($\mu$m) & ($\%$) & (s) & (S/N = 5) \\

\hline
3.05 ice MK       &  2.970	&	3.125 &	0.155	&	74	& 15  $\times$ 4 s &	13$^m$.0 \\
3.29 PAH MK 	    &  3.264	&	3.318	& 	0.054	&	70	& 	30  $\times$ 2 s & 11$^m$.9 \\
3.4 nbL              &  3.379	&	3.451	&	0.072	&	85	& 30  $\times$ 2 s &	12$^m$.2 \\
 3.5 mbL            & 3.383	&	3.594	&	0.211	&	82	& 30  $\times$ 2 s &	13$^m$.3 \\
3.6 nbL$'$           & 3.560	&	3.625	&	0.065	&	73 	& 30  $\times$ 2 s &	12$^m$.1 \\
3.99 (cont) 	  & 3.964	&	4.016 	&	0.052	&	80	& 30  $\times$ 2 s &	11$^m$.8  \\
\hline

\end{tabular}
    \end{center}
    \begin{flushleft}
\begin{footnotesize}
*  5$\sigma$ sensitivity thresholds were determined using the exposure time calculator for 9 $\times$ 1 min integration time. 
\end{footnotesize}
\end{flushleft}
\end{table*}

\section{Data Reduction}\label{Section3}

Data reduction was carried out using Starlink\footnote{http://starlink.eao.hawaii.edu/starlink} / ORAC-DR\footnote{http://www.starlink.ac.uk/docs/sun230.htx/sun230.html} Data Reduction pipeline software. A recipe was used to observe and reduce the data for each filter. This recipe takes an imaging observation comprising a series of jittered object frames and a dark frame, and a predetermined flat- frame, to make a calibrated, trimmed mosaic automatically. This recipe performs bad-pixel masking, null debiassing, dark subtraction, flat-field division, feature detection and matching between object frames, and resampling. The recipe makes the mosaic by applying offsets in intensity to give the most consistent result amongst the overlapping regions. For each cycle of jittered frames, the recipe creates a mosaic, which is then added into a master mosaic to improve the signal to noise. The final mosaic files were obtained by the combination of 2 sets of frames imaged with 9 different jitter positions. The mosaics were not trimmed to the dimensions of a single frame, thus the noise is greater in the peripheral areas which have received less total exposure time.

In order to correct the coordinate information on the mosaics, astrometric corrections were applied by using Karma\footnote{http://www.atnf.csiro.au/computing/software/karma/} packages (Kvis and Koords). Spitzer - GLIMPSE\footnote{https://irsa.ipac.caltech.edu/data/SPITZER/GLIMPSE/overview.html} cutout images were used to obtain the coordinates of the reference stars used for this purpose. 

\begin{figure}
\begin{center}
  \leavevmode
  \setlength\intextsep{0pt}
    \includegraphics[scale=0.45]{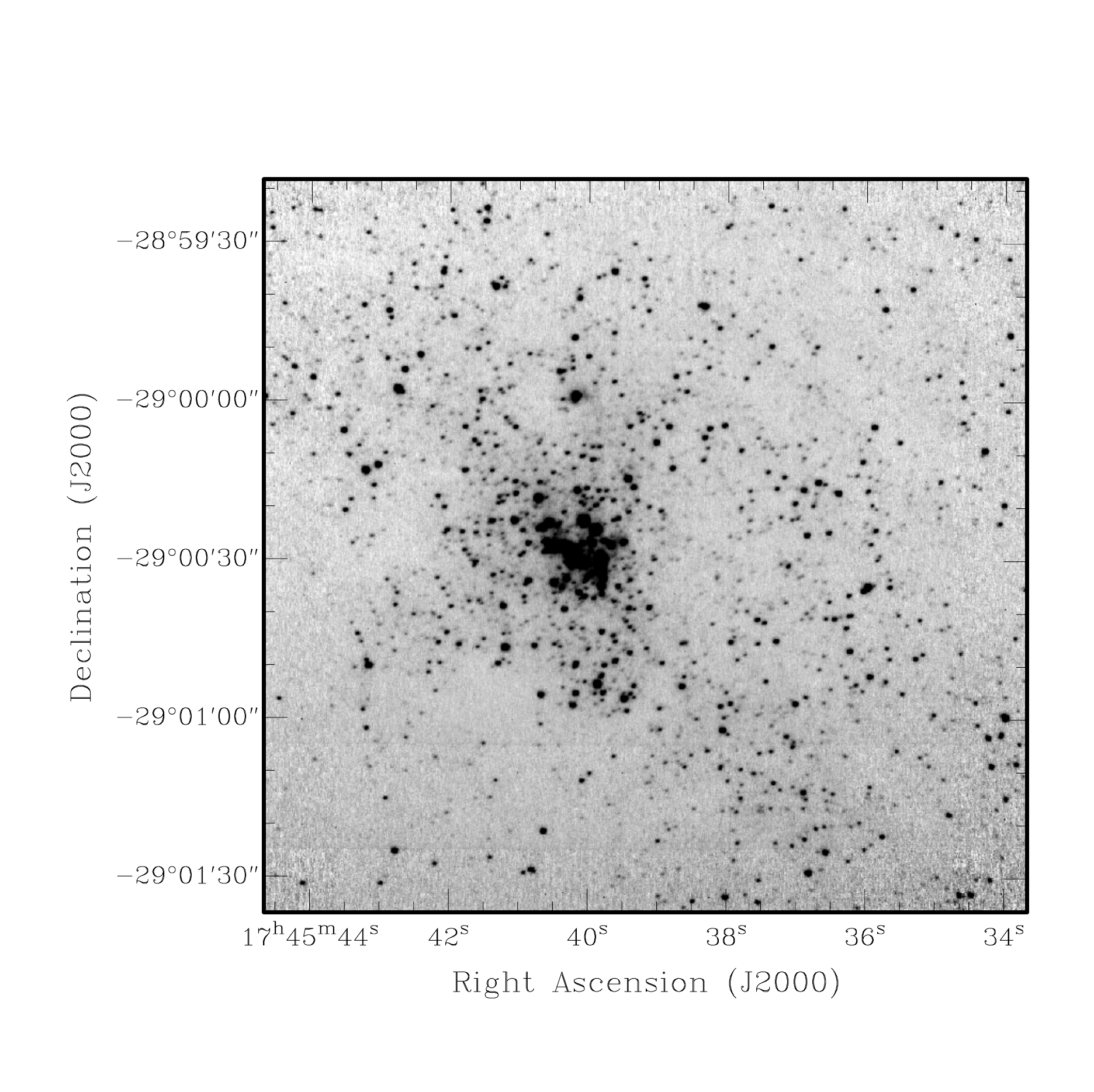}
       \caption{Image of the target field containing the Galactic Centre through the 3.4$\,\mu$m filter.} 
     \label{fig1}
     \end{center}
\end{figure}

The resultant image of the Galactic Centre obtained using the 3.4$\,\mu$m filter is presented in Figure \ref{fig1}.

\section{Data Analysis}\label{Section4}

\subsection{Photometric Measurements}

Photometric measurements of the field stars were carried out by using the Starlink / GAIA package\footnote{http://star-www.dur.ac.uk/~pdraper/gaia/gaia.html}. We used optimal photometry, which involves profile fitting of bright and isolated stars \citep{Naylor1998}. By examining different aperture sizes, the brightness variations were investigated and the best clipping radius was determined as 8.5 pixels (1.0 arcsec). This value is also consistent with the optimum aperture size indicated in \cite{Naylor1998}, as it is around 1.2--1.5 times larger than the FWHM ($\sim$ 5.5--6.5 pixels, 0.7--0.8 arc sec) of the point spread function of the stars. We measured brightnesses of 200 sources in the Galactic Centre. 

The measured brightness in magnitudes, $m$, were then converted into fluxes (W\,cm$^{-2}$ $\mu$m$^{-1}$) for each narrow-band filter using the equation below: 

\begin{equation} \label{eq1}
\Phi_{\lambda} = 10^{-0.4m} \times C_{\lambda} \times 10^{-15}  \, \rm{W\,cm^{-2} \mu m^{-1}}
\end{equation}

$C_{\lambda}$ is a scaling factor for each narrow-band filter, obtained by interpolating a linear fit through the zero point calibration values for the L and $L'$ broad-band filters of UKIRT ($C_{\lambda}$ = 7.31 and 5.24 at 3.45$\,\mu$m and 3.8$\,\mu$m as tabulated by the Spitzer Science Center magnitude--flux density converter\footnote{http://ssc.spitzer.caltech.edu/warmmission/propkit/pet/magtojy/}). We used $C_{\lambda}$ = 9.68, 8.20, 7.61, 7.01 and 6.42 at 3.05$\,\mu$m, 3.3$\,\mu$m, 3.4$\,\mu$m, 3.5$\,\mu$m and 3.6$\,\mu$m, respectively, for the conversation of the photometric measurements into the fluxes.

\subsection{Calibrating the Narrow Band Filters}

The measurement of fluxes through the narrow band filters provides low resolution spectra for every source in the field. To calibrate the counts obtained in these spectra we first need to determine the zero points for each of the filters. This is the flux level associated with 1 count measured by the instrument for each filter. We describe the technique used below, including how we overcame the issue of lack of calibrated standards, as measured through the narrow band filter set.

In the Galactic Centre, there are several well-studied bright L$-$band targets with the spectra available in the literature ([C02] and [M04]), e.g., GCIRS 1W, GCIRS 3, GCIRS 6E, GCIRS 7, GCIRS 9, GCIRS 13, GCIRS 16C, GCIRS 16NE, GCIRS 21, GCIRS 29. A group of these sources containing the prominent 3.4$\,\mu$m aliphatic feature in their spectra was employed to provide checks for the technique we discuss below.

\begin{figure}
	\begin{center}
  \leavevmode
\setlength\intextsep{0pt}
        \includegraphics[scale=0.45]{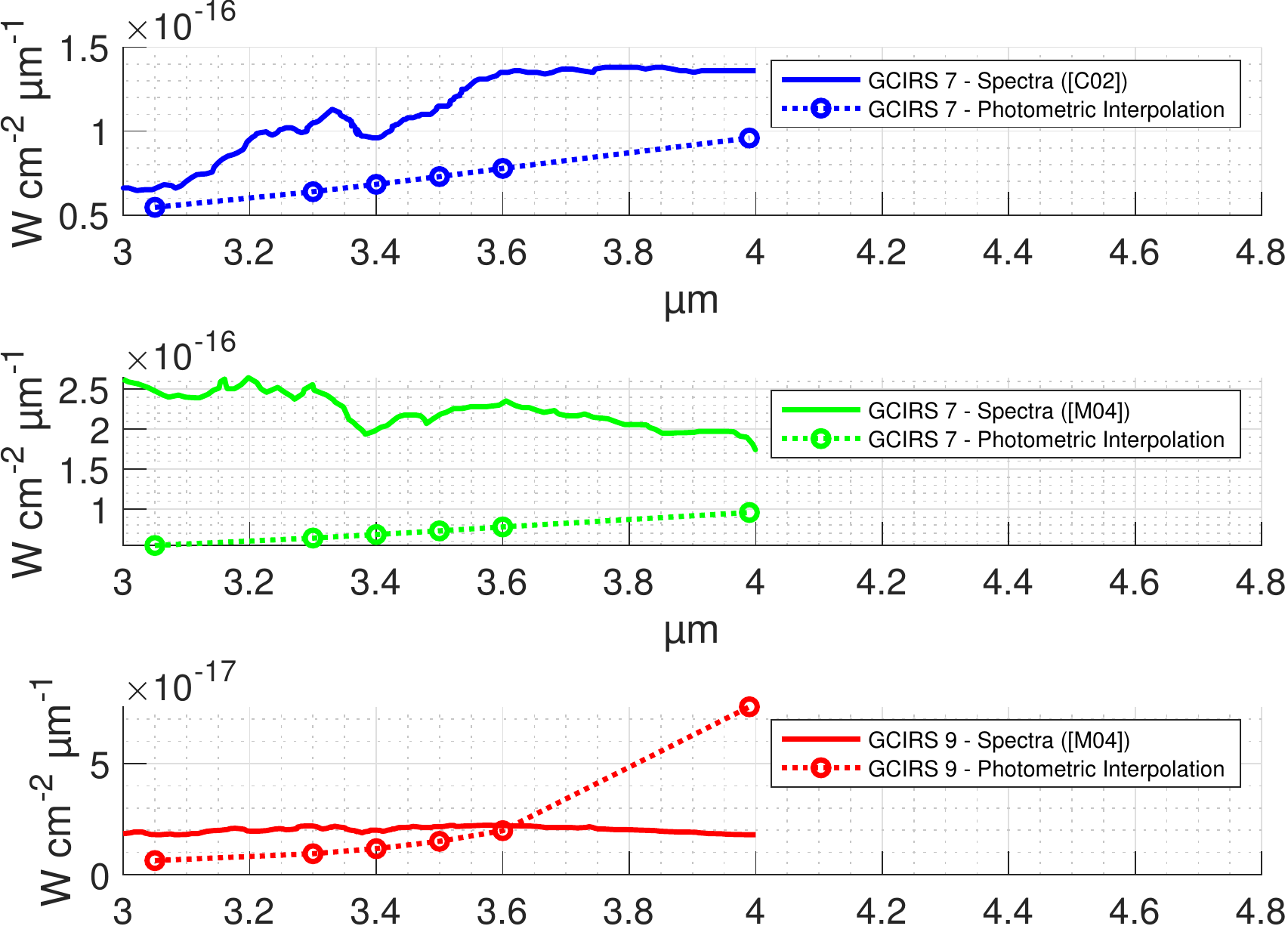}
       \caption{The low resolution spectra obtained by the interpolation of the photometric measurement with K - L - M filters are shown as smooth lines and compared with the spectra measured by others. Upper panel (blue): digitised spectra of GCIRS 7 ([C02]) plus photometric interpolation of GCIRS 7. Middle panel (green): digitised spectra of GCIRS 7 ([M04])  plus photometric interpolation of GCIRS 7.  Bottom panel (red): digitised spectra of GCIRS 9 ([M04]) plus photometric interpolation of GCIRS 9.} 
     \label{fig2}
     \end{center}
\end{figure}

\textit{Analysis of the photometric reference data}: We first tried using well-calibrated sources with photometric data within the Galactic Centre to determine the zero points. We used 2MASS and Spitzer measurements of GCIRS 7 and GCIRS 9 and applied a quadratic polynomial fit to the K - L - M broad band filter fluxes. We derived the corresponding fluxes for each of the narrow band filters and compared with the digitised published spectra ([C02] and [M04]) in Figure \ref{fig2}. When interpolation was applied it had the affect of smoothing over the 3.4$\,\mu$m absorption feature, so masking the very feature we aimed to measure. Fluxes for the reference stars have not been determined for the filter set we used, only for the broad band filters. So, unless a source in the field is known, a priori, not to have the aliphatic feature present, it cannot be used to provide a reliable calibration source when measuring the strength of this feature in other sources in the field. A different calibration method is required.

\begin{figure*}
  \leavevmode
  \centering
         \includegraphics[scale=0.8]{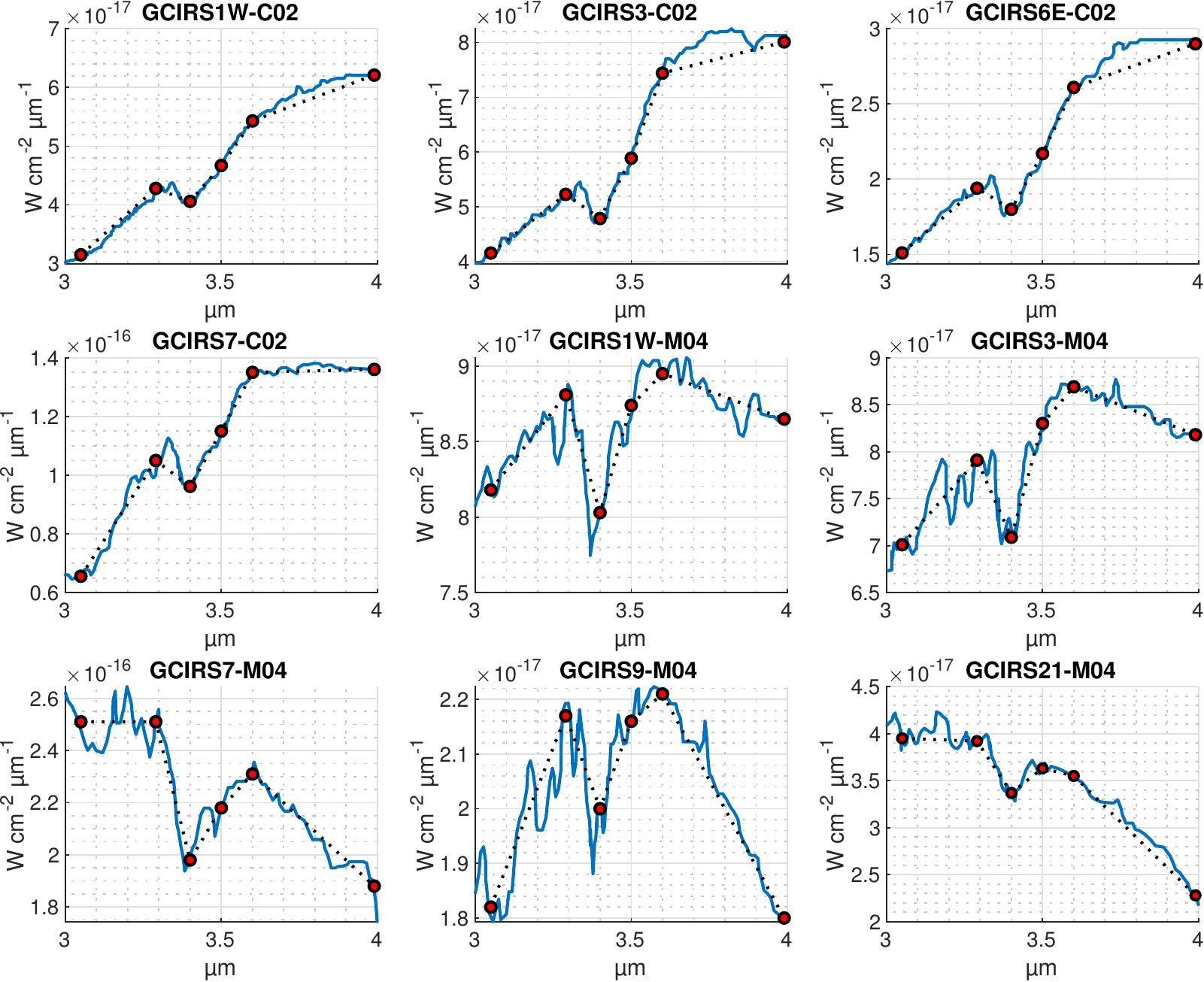}
       \caption{The spectra of the reference sources from [C02] and [M04] (blue lines). The corresponding fluxes for each filter at the related wavelengths (red dots) were used to calculate zero points, and so build up to low resolution spectra (black dots).} 
     \label{fig3}
\end{figure*}

\textit{Analysis of the spectroscopic reference data}: While we could then take the measured optical depth of the aliphatic feature for reference sources from the literature measurements, and use this to apply a correction to the photometric fluxes in order to indirectly yield optical depths for other sources in the field, instead, we chose another method based on spectral measurements of bright 3.4$\,\mu$m aliphatic absorption sources in the Galactic Centre: GCIRS 1W, GCIRS 3, GCIRS 6E, GCIRS 7, GCIRS 9 and GCIRS 21 ([C02], [M04]). Using their spectra, we estimated the fluxes through each of the narrow band filters and so obtained 9 sets of fluxes for the calibration: GCIRS1W-C02, GCIRS3-C02, GCIRS6E-C02, GCIRS7-C02, GCIRS1W-M04, GCIRS3-M04, GCIRS7-M04, GCIRS9-M04 and GCIRS21-M04 (note that three sources were studied by both [C02] and [M04]). We used these to provide zero points based on the number of counts measured through the filters. The flux sets are listed in Table \ref{tab:3}, and corresponding zero point sets are given in Table \ref{tab:4}. The spectra of the reference sources from [C02] and of the resultant low resolution spectra used for calibration are shown in Figure \ref{fig3}. 

\begin{table*}
 \begin{center}
  \caption{The fluxes (W cm$^{-2}$ $\mu$m$^{-1}$) obtained based on the spectra of the reference sources from the literature.}
  \label{tab:3}
   \centering

   \begin{tabular}{| c | c | c | >{\centering\arraybackslash}p{1.6 cm}| >{\centering\arraybackslash}p{1.9 cm} | >{\centering\arraybackslash}p{1.5 cm} | >{\centering\arraybackslash}p{1.4 cm} | >{\centering\arraybackslash}p{1.4 cm} | >{\centering\arraybackslash}p{1.5 cm} |}

       \hline
\multicolumn{2}{| c |}{\multirow{2}{*}{Sources}}&\multirow{2}{*}{Flux Sets}&\multicolumn{6}{c|}{Fluxes ($\times$\,10$^{-17}$ W cm$^{-2}$ $\mu$m$^{-1}$)}\\
  \cline{4-9}
  
\multicolumn{2}{|c|}{}& & 3.05 Ice MK 	& 3.29 PAH MK 	& 3.4 nbL & 	3.5 mbL & 	3.6 nbL	& 	3.99 (cont)  \\
  \hline
&GCIRS 1W  	&	GCIRS1W-C02 	&	3.15	&	4.28	&	4.06	&	4.67	&	5.43	&	6.21	\\

\multirow{1}{*}{}&	GCIRS 3 &	GCIRS3-C02	&	4.16	&	5.23	&	4.79	&	5.89	&	7.44	&	8.01	\\

[C02] &GCIRS 6E	&	GCIRS6E-C02 	&	1.51	&	1.94	&	1.80	&	2.17	&	2.61	&	2.90	\\

&GCIRS 7  &	GCIRS7-C02 	&	6.56	&	10.51	&	9.62	&	11.49	&	13.53	&	13.60	\\
  \hline
&GCIRS 1W & GCIRS1W-M04 	&	8.18	&	8.81	&	8.03	&	8.74	&	8.95	&	8.65	\\

&GCIRS 3	& GCIRS3-M04	 	&	7.01	&	7.91	&	7.09	&	8.30	&	8.69	&	8.18	\\

\multirow{1}{*}{}&GCIRS 7	& GCIRS7-M04 		&	25.14	&	25.14	&	19.79	&	21.77	&	23.13	&	18.77	\\

[M04]&GCIRS 9	& GCIRS9-M04 		&	1.82	&	2.17	&	2.00	&	2.16	&	2.21	&	1.80	\\

&GCIRS 21	 & 	GCIRS21-M04	 &	3.95	&	3.92	&	3.37	&	3.63	&	3.55	&	2.28	\\

   \hline
\end{tabular}
\end{center}
\end{table*}

 \begin{table*}
 \begin{center}
  \caption{The zero points (W cm$^{-2}$ $\mu$m$^{-1}$) calculated by using the fluxes obtained based on the 9 reference spectra.}
  \label{tab:4}
  \centering  
  \begin{tabular}{| c | c | c | >{\centering\arraybackslash}p{1.6 cm}| >{\centering\arraybackslash}p{1.7 cm} | >{\centering\arraybackslash}p{1.5 cm} | >{\centering\arraybackslash}p{1.4 cm} | >{\centering\arraybackslash}p{1.4 cm} | }

       \hline
\multicolumn{2}{| c |}{\multirow{2}{*}{Sources}}&\multirow{2}{*}{Zero Point Sets}&\multicolumn{5}{c|}{Zero Points ($\times$\,10$^{-8}$ W cm$^{-2}$ $\mu$m$^{-1}$)}\\
  \cline{4-8}
  
\multicolumn{2}{|c|}{}& & 3.05 Ice MK & 3.3 PAH MK & 3.4 nbL & 3.5 mbL & 3.6 nbL\\
  \hline
&GCIRS 1W  & GCIRS1W-C02 &	1.05	&	5.28	&	2.30	&	0.74	&	2.44	\\

\multirow{1}{*}{}&GCIRS 3 & GCIRS3-C02 &	2.65	&	9.03	&	3.46	&	1.10	&	3.53	\\

[C02] &GCIRS 6E	&GCIRS6E-C02 &	1.61	&	7.94	&	3.67	&	1.18	&	4.11	\\

&GCIRS 7  &GCIRS7-C02 &	1.38	&	7.03	&	3.12	&	1.00	&	3.34	\\

&GCIRS 1W & GCIRS1W-M04 &	2.74	&	10.86	&	4.56	&	1.38	&	4.02	\\

\multirow{1}{*}{}&GCIRS 3	& GCIRS3-M04 &	4.46	&	13.65	&	5.12	&	1.55	&	4.13	\\

[M04]&GCIRS 7	& GCIRS7-M04 &	5.27	&	16.82	&	6.43	&	1.90	&	5.71	\\

 &GCIRS 9	& GCIRS9-M04 	&	3.27	&	12.54	&	5.30	&	1.58	&	4.62	\\
 
&GCIRS 21	 & GCIRS21-M04 &	4.17	&	12.42	&	4.52	&	1.31	&	3.48	\\

   \hline
\end{tabular}
\end{center}
\end{table*}

While spectroscopic observations cannot provide accurate absolute flux levels, the relative fluxes between the filters are reliable, as is needed when determining the 3.4$\,\mu$m optical depth. The absolute flux levels obtained using this method were in fact found to be 2 - 3 times higher than the value when the direct photometric calibration of the broad band fluxes was used (see Figure \ref{fig2}).

\textit{Analysis of the resultant optical thicknesses}: We used the resultant zero points (Table \ref{tab:4}) to calibrate the spectrophotometric data to obtain low resolution spectra. For each of these 9 calibration sets we then flux calibrated the data for a group of test sources: GCIRS 1W, GCIRS 3, GCIRS 7, GCIRS 9, GCIRS 13, GCIRS 16C, GCIRS 16NE, GCIRS 21 and GCIRS 29. These stars were selected as they all have optical depths reported in [M04], to compare our own determinations of optical depth to. We preferred bright stars (m$_{L}$ $<$ 9$^{m}$) but avoided the stars in the most crowded regions of the field.

\begin{figure}
  \leavevmode
  \centering
         \includegraphics[scale=0.45]{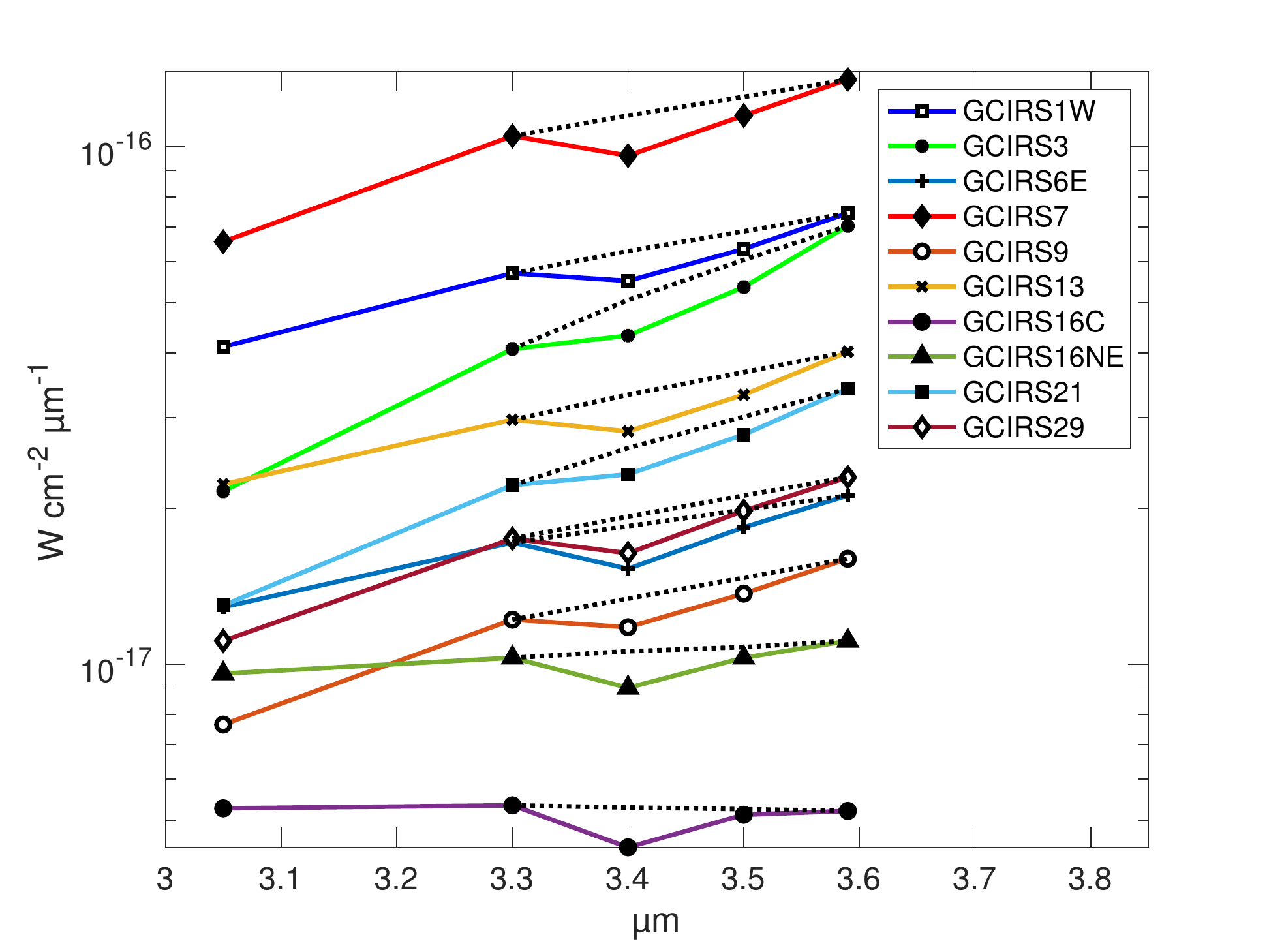}
       \caption{The calibrated fluxes were combined to obtain low-resolution calibrated spectra of the test sources. The 3.4$\,\mu$m optical depths were calculated by interpolating across the aliphatic absorption feature from 3.3$\,\mu$m  to 3.6$\,\mu$m to yield a linear continuum. This figure shows spectra calibrated using the GCISR7-C02 flux set.} 
     \label{fig4}
\end{figure}

\begin{figure}
  \leavevmode
  \centering
          \includegraphics[scale=0.45]{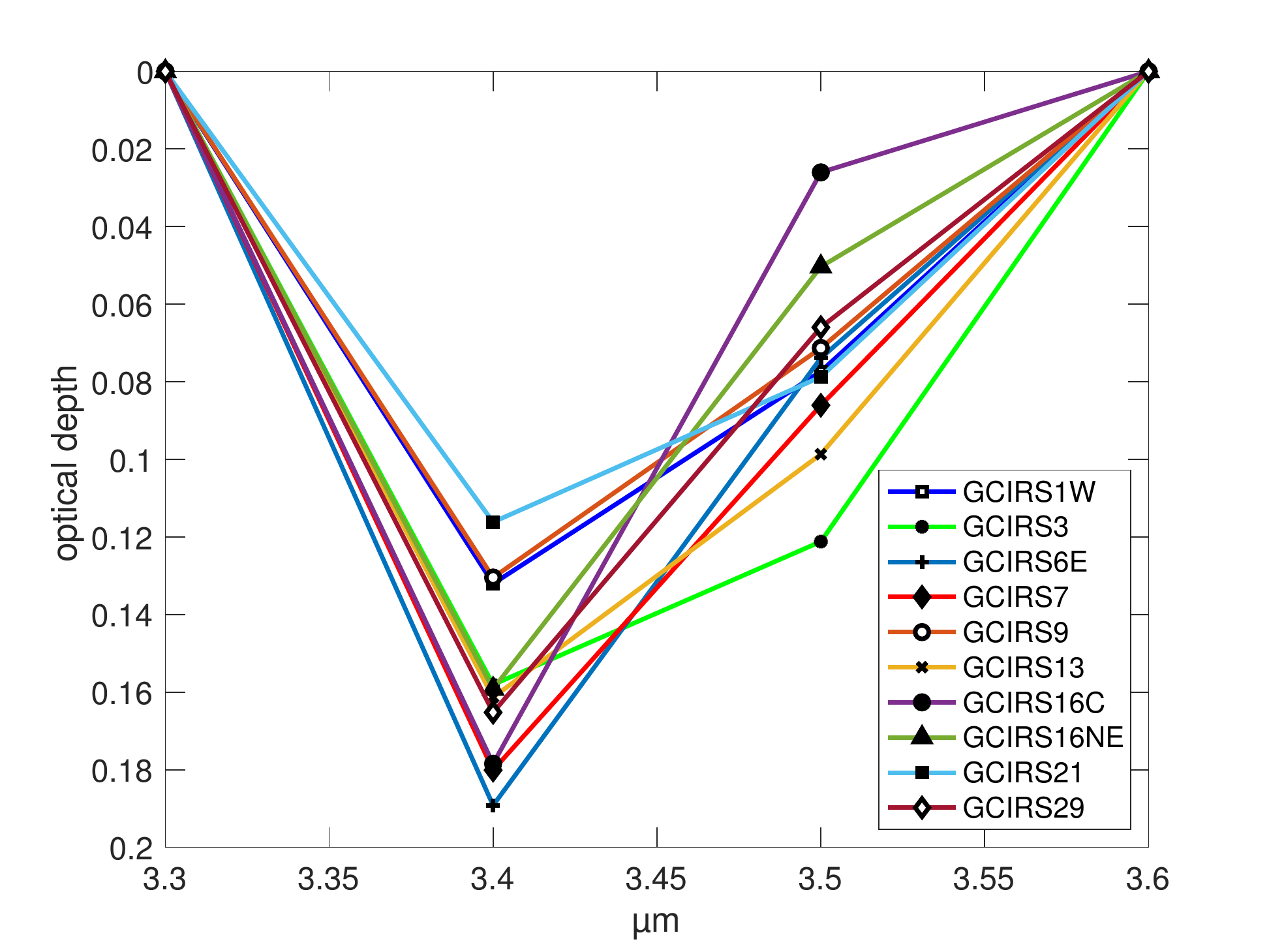}
                 \caption{The 3.4$\,\mu$m absorption feature optical depth derived from the data shown in Figure \ref{fig4}.} 
     \label{fig5}
\end{figure}

Since I/I$_{0}$ = $\exp(-\tau)$ then the optical depth is given by $\tau = -ln$(I/I$_{0}$), where I is the measured flux and I$_{0}$ is the interpolated flux. Thus the optical depths are calculated by using the fluxes measured by the 3.4$\,\mu$m filter and the fluxes calculated by interpolating across the aliphatic absorption feature from 3.3$\,\mu$m to 3.6$\,\mu$m to yield a linear continuum. Figure \ref{fig4} shows an example for calibrated spectra by GCIRS7-C02 and the corresponding continuum. Figure \ref{fig5} shows 3.4$\,\mu$m optical depth obtained from these spectra. We note that there is also a small optical depth derived at 3.5$\,\mu$m following the interpolation between 3.3$\,\mu$m and 3.6$\,\mu$m. However, it is much smaller than the optical depth derived at 3.4$\,\mu$m, generally about one-third the value.

We list the resultant 3.4$\,\mu$m optical depth values in Table \ref{tab:5}, together with the values reported by [M04] from their full-resolution spectra. While this method produces a simplification of the spectra, it is clear that the low resolution spectra still yield reasonable values for the optical depth. The difference in optical depth derived between the low and full-resolution spectra is less than the difference in optical depths reported by [C02] and [M04] for GCIRS 7.

 \begin{table*}
 \begin{center}
 \footnotesize
  \caption{The 3.4 $\mu$m optical depths for the test sources based on 9 different calibration sets. Results are compared with the reported values in the literature in the bottom row.}
  \label{tab:5}
  \centering  
   \begin{tabular}{| c | >{\centering\arraybackslash}p{1.1 cm} | >{\centering\arraybackslash}p{1.0 cm}  |>{\centering\arraybackslash}p{1.0 cm} |>{\centering\arraybackslash}p{1.0 cm}  | >{\centering\arraybackslash}p{1.0 cm}  | >{\centering\arraybackslash}p{1.1 cm} | >{\centering\arraybackslash}p{1.4 cm}  | >{\centering\arraybackslash}p{1.0 cm} |>{\centering\arraybackslash}p{1.0 cm} | }

 \hline
Zero Point   & \multicolumn{9}{c|}{3.4 $\mu$m Optical Depth ($\tau_{3.4\,\mu m}$)}  \\    
\cline{2-10}		
 		Calibration Sets&IRS 1W&	IRS 3	& IRS 7	&IRS 9	&IRS 13&	IRS 16C&	IRS 16NE&	IRS 21& IRS 29\\
\hline
GCIRS1W-C02	&	0.14	&	0.16	&	0.19	&	0.14	&	0.17	&	0.19	&	0.17	&	0.12	&	0.17\\
GCIRS3-C02	&	0.21	&	0.22	&	0.26	&	0.21	&	0.23	&	0.27	&	0.24	&	0.18	&	0.24\\
GCIRS6E-C02	&	0.13	&	0.16	&	0.17	&	0.13	&	0.16	&	0.17	&	0.15	&	0.11	&	0.16\\
GCIRS7-C02	&	0.13	&	0.16	&	0.18	&	0.13	&	0.16	&	0.18	&	0.16	&	0.12	&	0.17\\
GCIRS1W-M04	&	0.10	&	0.11	&	0.15	&	0.10	&	0.12	&	0.16	&	0.14	&	0.07	&	0.13\\
GCIRS3-M04	&	0.12	&	0.12	&	0.17	&	0.12	&	0.15	&	0.20	&	0.17	&	0.09	&	0.15\\
GCIRS7-M04	&	0.16	&	0.16	&	0.21	&	0.16	&	0.18	&	0.22	&	0.20	&	0.13	&	0.19\\
GCIRS9-M04	&	0.12	&	0.13	&	0.17	&	0.12	&	0.15	&	0.18	&	0.16	&	0.09	&	0.15\\
GCIRS21-M04	&	0.14	&	0.13	&	0.19	&	0.14	&	0.17	&	0.22	&	0.19	&	0.11	&	0.17\\
\hline
[M04] &  	0.16	&	0.20	&	0.20	&	0.16	&	0.16	&	0.14	&	0.14	&	0.15	&	0.27\\
\hline
\end{tabular}
\end{center}
\end{table*}

We list these differences in Table \ref{tab:6}. We then selected those which yielded the smallest average absolute mean differences ($\mid$$\Delta$$\mid$) as the best calibration options. Thus GCIRS1W-C02, GCIRS7-C02 and GCIRS7-M04 were chosen to calibrate all spectrophotometric data of Galactic Centre (See Table \ref{tab:6}).

\begin{table*}
 \begin{center}
  \caption{Difference between derived optical depths from their values reported in the literature and the average of the absolute mean differences.} 
\centering
  \label{tab:6}
  \begin{tabular}{| c | c | c | c | c | c | c | c | c | c | c |}
    \hline
Zero Point   & \multicolumn{9}{c|}{Differences - $\Delta$} & \multirow{2}{*}{Average $\mid$$\Delta$$\mid$}  \\   
	 \cline{2-10}		
Calibration Sets &IRS1W&	IRS3	&IRS7&	IRS9&	IRS13& IRS16C& 	IRS16NE& IRS21	& IRS29	 & \\

\hline
GCIRS1W-C02	&	0.02	&	0.04	&	0.01	&	0.02	&	-0.01	&	-0.05	&	-0.03	&	0.03	&	0.10 	&	0.03\\ 
GCIRS3-C02	&	-0.05	&	-0.02	&	-0.06	&	-0.05	&	-0.07	&	-0.13	&	-0.10	&	-0.03	&	0.03 	&	0.06\\ 
GCIRS6E-C02	&	0.03	&	0.04	&	0.03	&	0.03	&	0.00	&	-0.03	&	-0.01	&	0.04	&	0.11 	&	0.04\\ 
GCIRS7-C02	&	0.03	&	0.04	&	0.02	&	0.03	&	0.00	&	-0.04	&	-0.02	&	0.03	&	0.10	&	0.03\\ 
GCIRS1W-M04	&	0.06	&	0.09	&	0.05	&	0.06	&	0.04	&	-0.02	&	0.00	&	0.08	&	0.14 	&	0.06\\ 
GCIRS3-M04	&	0.04	&	0.08	&	0.03	&	0.04	&	0.01	&	-0.06	&	-0.03	&	0.06	&	0.12	&	0.05\\ 
GCIRS7-M04	&	0.00	&	0.04	&	-0.01	&	0.00	&	-0.02	&	-0.08	&	-0.06	&	0.02	&	0.08	&	0.03\\ 
GCIRS9-M04	&	0.04	&	0.07	&	0.03	&	0.04	&	0.01	&	-0.04	&	-0.02	&	0.06	&	0.12	&	0.05\\ 
GCIRS21-M04	&	0.02	&	0.07	&	0.01	&	0.02	&	-0.01	&	-0.08	&	-0.05	&	0.04	&	0.10	&	0.04 \\ 

\hline
\end{tabular}
    \end{center}
  \end{table*}

Applying the calibration factors resulting from these three sources, we then calibrated the data set gathered for the Galactic Centre field. We determined the brightest 200 sources with fluxes of $m_{3.6}$ $<$ 10$^{m}$ and then derived their 3.4$\,\mu$m optical depths.

We then checked for consistency in the derived optical depths among these three different calibration sources to ensure the efficacy of this method. First, we found that the values of the optical depths for the sources in the field changed relatively little across the field. Using GCIRS1W-C02, GCIRS7-C02 and GCIRS7-M04 as the calibration sources, the mean $\tau_{3.4\,\mu m}$ value was found as 0.207$\pm$0.063, 0.201$\pm$0.063 and 0.237$\pm$0.067, respectively. Given this consistency, we then averaged the values from these three calibration sources, and obtained an average optical depth of $\tau_{3.4\,\mu m}$ = 0.215$\pm$0.064 across the sources in the field of Galactic Centre. Furthermore, depending on the choice of the calibration source, the difference in the optical depth derived for each source is also consistent. For instance, the average $\tau_{3.4\,\mu m}$ is 0.007$\pm$0.001 higher using GCIRS1W-C02 cf. GCIRS 7-C02, 0.036$\pm$0.014 higher using GCIRS7-M04 cf. GCIRS7-C02, and 0.030$\pm$0.013 higher using GCIRS7-M04 cf. GCIRS1W-C02.

GCIRS 7 is the brightest source and it has higher resolution (R=1200) in the spectrum provided by [C02]. Thus, in order to calculate optical depths for each source, we chose GCIRS7-C02 to calibrate the field.

 \section{Results}	\label{Section5}

We present the fluxes at 3.3$\,\mu$m, 3.4$\,\mu$m and 3.6$\,\mu$m for all the sources measured in the field in Table \ref{tab:7}, calculated using GCIRS7-C02 as the calibration source, as described in the previous section. In Table \ref{tab:8} we list the optical depth at 3.4 $\,\mu$m derived from the calibrated fluxes, together with the celestial coordinates of the related sources. For comparison, we also noted in the footnote of Table \ref{tab:7} and Table \ref{tab:8} the corresponding IDs of the calibration sources and test group of stars so that they can be identified.

\begin{table*}
 \begin{center}
  \caption{The calibrated fluxes ($\times$\,10$^{-18}$ W\,cm$^{-2}$ $\mu$m$^{-1}$) of the Galactic Centre sources used for optical depth calculations.} 
\centering
  \label{tab:7}
  \begin{tabular}{|  p {0.6 cm}  p {0.7 cm} p {0.7 cm} p {0.7 cm} | p {0.6 cm}  p {0.7 cm} p {0.7 cm} p {0.7 cm}  | p {0.6 cm}  p {0.7 cm} p {0.7 cm} p {0.7 cm} | p {0.6 cm}  p {0.7 cm} p {0.7 cm} p {0.7 cm} |}
    \hline

Source &  \multicolumn{3}{c|}{Fluxes } & Source & \multicolumn{3}{c|}{Fluxes } & Source & \multicolumn{3}{c|}{Fluxes}  &  Source &\multicolumn{3}{c|}{Fluxes} \\

\multirow{1}{*}{No} &	3.3$\,\mu$m & 	3.4$\,\mu$m 	&	3.6$\,\mu$m	&	\multirow{1}{*}{No} &	3.3$\,\mu$m & 	3.4$\,\mu$m	&3.6$\,\mu$m	 &\multirow{1}{*}{No} &	3.3$\,\mu$m & 	3.4$\,\mu$m 	&	3.6$\,\mu$m	& \multirow{1}{*}{No} &	3.3$\,\mu$m & 	3.4$\,\mu$m 	&	3.6$\,\mu$m	\\
    \hline
1	&	105.3	&	96.3	&	135.5	&	51	&	6.9	&	6.0	&	8.1	&	101	&	5.3	&	4.5	&	4.9	&	151	&	2.7	&	2.6	&	2.9	\\
2	&	57.1	&	55.2	&	74.6	&	52	&	8.5	&	6.1	&	8.0	&	102	&	4.0	&	3.7	&	4.8	&	152	&	2.5	&	2.1	&	2.9	\\
3	&	40.8	&	43.3	&	70.5	&	53	&	7.7	&	5.7	&	7.9	&	103	&	4.0	&	3.2	&	4.7	&	153	&	2.3	&	1.9	&	2.9	\\
4	&	29.7	&	28.3	&	40.2	&	54	&	5.3	&	5.1	&	7.9	&	104	&	3.8	&	3.3	&	4.6	&	154	&	1.9	&	1.7	&	2.8	\\
5	&	28.4	&	25.4	&	34.9	&	55	&	8.3	&	6.4	&	7.8	&	105	&	3.2	&	3.1	&	4.5	&	155	&	2.5	&	2.1	&	2.8	\\
6	&	22.2	&	23.3	&	34.1	&	56	&	5.2	&	5.0	&	7.8	&	106	&	4.0	&	3.3	&	4.5	&	156	&	2.5	&	2.0	&	2.8	\\
7	&	23.4	&	22.8	&	33.3	&	57	&	6.2	&	5.9	&	7.8	&	107	&	3.0	&	2.7	&	4.5	&	157	&	2.2	&	1.9	&	2.8	\\
8	&	20.4	&	19.0	&	32.3	&	58	&	5.6	&	5.6	&	7.6	&	108	&	3.1	&	2.9	&	4.4	&	158	&	2.2	&	1.9	&	2.8	\\
9	&	20.9	&	20.2	&	28.1	&	59	&	5.9	&	5.2	&	7.5	&	109	&	3.9	&	3.1	&	4.4	&	159	&	2.3	&	2.1	&	2.8	\\
10	&	19.5	&	18.2	&	25.8	&	60	&	5.3	&	4.9	&	7.5	&	110	&	3.9	&	3.7	&	4.4	&	160	&	2.4	&	2.0	&	2.7	\\
11	&	22.6	&	21.4	&	24.1	&	61	&	7.3	&	5.5	&	7.4	&	111	&	4.4	&	3.5	&	4.4	&	161	&	2.2	&	1.9	&	2.7	\\
12	&	15.3	&	15.1	&	24.1	&	62	&	8.3	&	6.3	&	7.3	&	112	&	2.3	&	2.4	&	4.3	&	162	&	2.4	&	2.0	&	2.7	\\
13	&	17.5	&	16.4	&	23.1	&	63	&	5.3	&	5.4	&	7.2	&	113	&	3.1	&	3.0	&	4.3	&	163	&	2.3	&	1.8	&	2.7	\\
14	&	17.2	&	15.3	&	21.2	&	64	&	6.1	&	4.8	&	7.2	&	114	&	4.1	&	3.4	&	4.2	&	164	&	2.9	&	2.4	&	2.6	\\
15	&	12.7	&	13.7	&	21.0	&	65	&	5.5	&	5.4	&	7.1	&	115	&	3.9	&	3.2	&	4.2	&	165	&	2.1	&	1.8	&	2.6	\\
16	&	17.7	&	14.8	&	19.8	&	66	&	5.2	&	4.6	&	7.1	&	116	&	3.3	&	2.8	&	4.2	&	166	&	2.8	&	2.1	&	2.6	\\
17	&	17.8	&	15.1	&	18.5	&	67	&	7.2	&	5.6	&	7.1	&	117	&	3.6	&	3.0	&	4.0	&	167	&	2.5	&	2.3	&	2.6	\\
18	&	12.5	&	11.3	&	16.4	&	68	&	6.0	&	5.0	&	7.0	&	118	&	3.9	&	2.7	&	3.9	&	168	&	1.6	&	1.6	&	2.6	\\
19	&	12.2	&	11.9	&	16.2	&	69	&	5.4	&	4.8	&	6.7	&	119	&	4.0	&	3.4	&	3.9	&	169	&	2.4	&	1.9	&	2.5	\\
20	&	10.9	&	11.1	&	16.0	&	70	&	6.4	&	5.4	&	6.6	&	120	&	3.3	&	2.8	&	3.9	&	170	&	2.9	&	2.3	&	2.5	\\
21	&	12.2	&	11.8	&	16.0	&	71	&	5.9	&	5.0	&	6.5	&	121	&	3.4	&	3.0	&	3.8	&	171	&	1.8	&	1.5	&	2.4	\\
22	&	10.5	&	10.5	&	15.4	&	72	&	4.6	&	4.2	&	6.5	&	122	&	4.4	&	2.9	&	3.8	&	172	&	2.7	&	2.1	&	2.4	\\
23	&	13.1	&	11.7	&	15.2	&	73	&	6.4	&	5.3	&	6.5	&	123	&	3.0	&	3.0	&	3.8	&	173	&	2.5	&	1.8	&	2.3	\\
24	&	6.8	&	7.8	&	14.9	&	74	&	5.4	&	4.6	&	6.4	&	124	&	3.7	&	3.1	&	3.7	&	174	&	2.2	&	1.8	&	2.3	\\
25	&	13.9	&	11.7	&	14.6	&	75	&	6.8	&	6.2	&	6.3	&	125	&	2.6	&	2.1	&	3.7	&	175	&	1.9	&	1.7	&	2.3	\\
26	&	12.1	&	10.1	&	13.7	&	76	&	5.7	&	4.8	&	6.3	&	126	&	3.1	&	2.5	&	3.7	&	176	&	1.8	&	1.5	&	2.3	\\
27	&	10.9	&	9.4	&	13.5	&	77	&	6.3	&	5.4	&	6.2	&	127	&	2.3	&	2.1	&	3.7	&	177	&	1.7	&	1.5	&	2.3	\\
28	&	8.3	&	8.1	&	12.6	&	78	&	5.2	&	4.7	&	6.2	&	128	&	3.0	&	2.9	&	3.6	&	178	&	2.5	&	2.2	&	2.3	\\
29	&	11.6	&	10.1	&	11.8	&	79	&	5.9	&	4.8	&	6.2	&	129	&	3.3	&	2.7	&	3.6	&	179	&	1.4	&	1.5	&	2.2	\\
30	&	6.0	&	6.8	&	11.2	&	80	&	4.9	&	4.3	&	6.2	&	130	&	3.0	&	2.6	&	3.5	&	180	&	2.0	&	1.8	&	2.2	\\
31	&	10.3	&	9.0	&	11.1	&	81	&	6.8	&	5.8	&	6.0	&	131	&	3.5	&	2.9	&	3.5	&	181	&	2.3	&	1.7	&	2.2	\\
32	&	9.1	&	7.4	&	10.8	&	82	&	5.3	&	4.5	&	6.0	&	132	&	3.5	&	2.5	&	3.5	&	182	&	1.7	&	1.6	&	2.1	\\
33	&	7.8	&	6.6	&	10.1	&	83	&	5.9	&	4.7	&	6.0	&	133	&	2.5	&	2.2	&	3.4	&	183	&	1.5	&	1.3	&	2.0	\\
34	&	8.6	&	8.0	&	9.9	&	84	&	4.5	&	4.2	&	5.9	&	134	&	3.2	&	2.7	&	3.4	&	184	&	1.2	&	1.3	&	2.0	\\
35	&	7.2	&	6.7	&	9.8	&	85	&	5.1	&	4.3	&	5.9	&	135	&	3.1	&	2.5	&	3.4	&	185	&	1.7	&	1.4	&	1.9	\\
36	&	9.8	&	8.3	&	9.8	&	86	&	6.3	&	4.4	&	5.9	&	136	&	2.7	&	2.4	&	3.4	&	186	&	1.6	&	1.4	&	1.9	\\
37	&	8.5	&	7.4	&	9.8	&	87	&	3.1	&	3.5	&	5.8	&	137	&	3.3	&	2.6	&	3.4	&	187	&	1.9	&	1.6	&	1.9	\\
38	&	6.0	&	5.7	&	9.7	&	88	&	4.1	&	3.9	&	5.6	&	138	&	1.7	&	1.8	&	3.3	&	188	&	2.3	&	1.7	&	1.8	\\
39	&	6.0	&	5.7	&	9.7	&	89	&	4.2	&	3.6	&	5.6	&	139	&	2.1	&	2.1	&	3.3	&	189	&	1.7	&	1.3	&	1.8	\\
40	&	6.0	&	5.6	&	9.4	&	90	&	5.0	&	4.9	&	5.6	&	140	&	3.6	&	3.2	&	3.2	&	190	&	1.3	&	1.1	&	1.8	\\
41	&	6.3	&	6.4	&	9.3	&	91	&	3.8	&	3.6	&	5.6	&	141	&	2.6	&	2.4	&	3.2	&	191	&	1.6	&	1.4	&	1.7	\\
42	&	9.4	&	6.1	&	9.2	&	92	&	5.2	&	4.1	&	5.5	&	142	&	2.9	&	2.1	&	3.1	&	192	&	1.6	&	1.3	&	1.7	\\
43	&	4.3	&	4.9	&	9.0	&	93	&	4.1	&	3.8	&	5.5	&	143	&	2.5	&	2.1	&	3.1	&	193	&	1.5	&	1.3	&	1.7	\\
44	&	5.2	&	5.3	&	8.8	&	94	&	3.9	&	3.6	&	5.3	&	144	&	2.6	&	2.2	&	3.0	&	194	&	2.2	&	1.4	&	1.7	\\
45	&	8.5	&	7.1	&	8.7	&	95	&	6.0	&	3.9	&	5.3	&	145	&	2.1	&	2.1	&	3.0	&	195	&	1.5	&	1.3	&	1.6	\\
46	&	4.8	&	5.2	&	8.6	&	96	&	5.3	&	4.4	&	5.2	&	146	&	2.5	&	2.2	&	3.0	&	196	&	1.4	&	1.2	&	1.6	\\
47	&	7.5	&	6.7	&	8.6	&	97	&	4.5	&	4.1	&	5.2	&	147	&	2.2	&	1.9	&	3.0	&	197	&	1.1	&	1.2	&	1.6	\\
48	&	7.7	&	6.6	&	8.3	&	98	&	4.1	&	3.4	&	5.1	&	148	&	2.2	&	2.0	&	3.0	&	198	&	1.2	&	1.1	&	1.4	\\
49	&	7.0	&	6.4	&	8.3	&	99	&	3.4	&	3.1	&	5.0	&	149	&	2.4	&	2.2	&	3.0	&	199	&	1.0	&	0.9	&	1.2	\\
50	&	7.3	&	6.6	&	8.2	&	100	&	4.6	&	3.9	&	4.9	&	150	&	1.5	&	1.7	&	3.0	&	200	&	1.1	&	1.0	&	1.1	\\
\hline
\end{tabular}
  \begin{flushleft}
\begin{footnotesize}
The source numbers of the calibration sources and the test group stars: 1: GCIRS 7, 2: GCIRS 1W, 3: GCIRS 3, 4: GCIRS 13, 6: GCIRS 21, 13: GCIRS 29, 14: GCIRS 6E, 21: GCIRS 9, 31: GCIRS 16 NE and 96: GCIRS 16C.
\end{footnotesize}
\end{flushleft}

    \end{center}
  \end{table*}

\begin{table*}
 \begin{center}
  \caption{Celestial coordinates and the optical depths from the 3.4$\,\mu$m absorption feature along the line of sight of the Galactic Centre sources.} 
\centering
  \label{tab:8}
   \begin{tabular}{| p {0.3 cm}  p {0.9 cm} p {1.1 cm} p {0.4 cm} | p {0.3 cm}  p {0.9 cm} p {1.1 cm} p {0.4 cm}  | p {0.3 cm}  p {0.9 cm} p {1.1 cm} p {0.4 cm}  | p {0.3 cm}  p {0.9 cm} p {1.1 cm} p {0.4 cm} |}
    \hline

No & 	\textit {l}  & \textit {b} 	&	$\tau_{3.4}$	&	No  &	\textit {l}   & \textit {b} 	&	$\tau_{3.4}$	&  No &	\textit {l}  & \textit {b} 	&	$\tau_{3.4}$	&  No &	\textit {l}  & \textit {b} 	&	$\tau_{3.4}$	\\
    \hline

1	&	266.4168	&	-29.0062	&	0.18	&	51	&	266.3996	&	-29.0143	&	0.20	&	101	&	266.4115	&	-29.0034	&	0.13	&	151	&	266.4169	&	-29.0198	&	0.07	\\
2	&	266.4185	&	-29.0075	&	0.13	&	52	&	266.4106	&	-28.9859	&	0.30	&	102	&	266.4149	&	-29.0135	&	0.14	&	152	&	266.3997	&	-29.0046	&	0.20	\\
3	&	266.4161	&	-29.0066	&	0.16	&	53	&	266.3912	&	-28.9964	&	0.31	&	103	&	266.3971	&	-28.9939	&	0.29	&	153	&	266.4150	&	-28.9963	&	0.24	\\
4	&	266.4158	&	-29.0081	&	0.16	&	54	&	266.4124	&	-29.0020	&	0.19	&	104	&	266.4000	&	-29.0030	&	0.20	&	154	&	266.3902	&	-29.0027	&	0.25	\\
5	&	266.4280	&	-28.9992	&	0.18	&	55	&	266.4227	&	-28.9904	&	0.25	&	105	&	266.4064	&	-29.0113	&	0.17	&	155	&	266.4343	&	-29.0006	&	0.21	\\
6	&	266.4175	&	-29.0084	&	0.12	&	56	&	266.4041	&	-29.0158	&	0.19	&	106	&	266.4165	&	-28.9975	&	0.22	&	156	&	266.3954	&	-28.9907	&	0.28	\\
7	&	266.4196	&	-29.0050	&	0.16	&	57	&	266.4171	&	-29.0099	&	0.12	&	107	&	266.3913	&	-29.0003	&	0.26	&	157	&	266.4269	&	-28.9939	&	0.26	\\
8	&	266.4346	&	-28.9871	&	0.25	&	58	&	266.4173	&	-29.0152	&	0.12	&	108	&	266.4017	&	-29.0113	&	0.18	&	158	&	266.3940	&	-29.0010	&	0.24	\\
9	&	266.4299	&	-29.0035	&	0.14	&	59	&	266.4022	&	-29.0236	&	0.22	&	109	&	266.4055	&	-28.9970	&	0.27	&	159	&	266.4304	&	-29.0121	&	0.16	\\
10	&	266.4189	&	-29.0062	&	0.17	&	60	&	266.4171	&	-28.9944	&	0.21	&	110	&	266.4202	&	-29.0073	&	0.11	&	160	&	266.4122	&	-29.0056	&	0.22	\\
11	&	266.4160	&	-29.0148	&	0.08	&	61	&	266.4227	&	-28.9897	&	0.28	&	111	&	266.3973	&	-29.0285	&	0.21	&	161	&	266.3976	&	-29.0152	&	0.23	\\
12	&	266.4173	&	-28.9996	&	0.19	&	62	&	266.3993	&	-29.0012	&	0.23	&	112	&	266.3978	&	-29.0020	&	0.21	&	162	&	266.4045	&	-29.0085	&	0.24	\\
13	&	266.4163	&	-29.0073	&	0.17	&	63	&	266.4198	&	-29.0127	&	0.11	&	113	&	266.4079	&	-29.0125	&	0.16	&	163	&	266.4012	&	-28.9897	&	0.30	\\
14	&	266.4152	&	-29.0074	&	0.19	&	64	&	266.3901	&	-28.9996	&	0.30	&	114	&	266.3968	&	-29.0134	&	0.18	&	164	&	266.4052	&	-29.0044	&	0.16	\\
15	&	266.4215	&	-29.0128	&	0.12	&	65	&	266.4180	&	-29.0108	&	0.12	&	115	&	266.4021	&	-29.0068	&	0.23	&	165	&	266.4129	&	-28.9951	&	0.22	\\
16	&	266.3914	&	-29.0165	&	0.22	&	66	&	266.4008	&	-29.0130	&	0.23	&	116	&	266.4056	&	-28.9896	&	0.25	&	166	&	266.3954	&	-29.0041	&	0.24	\\
17	&	266.3997	&	-29.0096	&	0.17	&	67	&	266.4015	&	-29.0047	&	0.23	&	117	&	266.4209	&	-29.0012	&	0.20	&	167	&	266.4298	&	-29.0171	&	0.11	\\
18	&	266.4221	&	-28.9938	&	0.20	&	68	&	266.3987	&	-28.9950	&	0.24	&	118	&	266.3983	&	-28.9879	&	0.35	&	168	&	266.3962	&	-28.9986	&	0.19	\\
19	&	266.4158	&	-29.0096	&	0.13	&	69	&	266.3925	&	-29.0175	&	0.19	&	119	&	266.4312	&	-29.0056	&	0.16	&	169	&	266.4247	&	-28.9898	&	0.25	\\
20	&	266.4144	&	-29.0155	&	0.13	&	70	&	266.4285	&	-28.9951	&	0.18	&	120	&	266.3993	&	-29.0066	&	0.21	&	170	&	266.3940	&	-28.9996	&	0.20	\\
21	&	266.4186	&	-29.0094	&	0.13	&	71	&	266.3923	&	-29.0190	&	0.19	&	121	&	266.3988	&	-29.0227	&	0.17	&	171	&	266.4348	&	-28.9947	&	0.25	\\
22	&	266.4297	&	-29.0138	&	0.14	&	72	&	266.4083	&	-29.0011	&	0.21	&	122	&	266.3922	&	-28.9859	&	0.39	&	172	&	266.3917	&	-28.9989	&	0.23	\\
23	&	266.4210	&	-29.0062	&	0.17	&	73	&	266.4138	&	-29.0044	&	0.18	&	123	&	266.4243	&	-29.0127	&	0.10	&	173	&	266.4028	&	-28.9970	&	0.32	\\
24	&	266.4040	&	-29.0269	&	0.20	&	74	&	266.3915	&	-29.0053	&	0.22	&	124	&	266.4132	&	-28.9934	&	0.18	&	174	&	266.4071	&	-28.9987	&	0.21	\\
25	&	266.4142	&	-29.0039	&	0.19	&	75	&	266.4174	&	-29.0159	&	0.08	&	125	&	266.4027	&	-28.9861	&	0.34	&	175	&	266.3957	&	-29.0163	&	0.19	\\
26	&	266.4096	&	-28.9949	&	0.23	&	76	&	266.4117	&	-29.0012	&	0.20	&	126	&	266.4087	&	-28.9960	&	0.27	&	176	&	266.4038	&	-28.9924	&	0.26	\\
27	&	266.4361	&	-28.9996	&	0.23	&	77	&	266.3935	&	-29.0258	&	0.16	&	127	&	266.4191	&	-28.9911	&	0.25	&	177	&	266.3908	&	-29.0123	&	0.25	\\
28	&	266.4084	&	-29.0172	&	0.19	&	78	&	266.4223	&	-29.0062	&	0.16	&	128	&	266.4336	&	-29.0253	&	0.10	&	178	&	266.4290	&	-29.0252	&	0.09	\\
29	&	266.4292	&	-29.0032	&	0.14	&	79	&	266.4331	&	-28.9986	&	0.22	&	129	&	266.4113	&	-28.9985	&	0.21	&	179	&	266.3980	&	-29.0182	&	0.10	\\
30	&	266.4194	&	-29.0153	&	0.12	&	80	&	266.4036	&	-29.0045	&	0.22	&	130	&	266.4087	&	-28.9872	&	0.22	&	180	&	266.4281	&	-29.0106	&	0.11	\\
31	&	266.4177	&	-29.0074	&	0.16	&	81	&	266.4083	&	-29.0259	&	0.11	&	131	&	266.4300	&	-28.9948	&	0.19	&	181	&	266.3894	&	-28.9901	&	0.29	\\
32	&	266.3927	&	-29.0024	&	0.26	&	82	&	266.4183	&	-28.9983	&	0.21	&	132	&	266.3933	&	-28.9913	&	0.34	&	182	&	266.4287	&	-29.0208	&	0.15	\\
33	&	266.4349	&	-28.9849	&	0.25	&	83	&	266.4034	&	-29.0069	&	0.24	&	133	&	266.3914	&	-29.0207	&	0.24	&	183	&	266.4213	&	-28.9979	&	0.23	\\
34	&	266.4281	&	-29.0235	&	0.12	&	84	&	266.4052	&	-29.0125	&	0.18	&	134	&	266.4163	&	-29.0029	&	0.20	&	184	&	266.4150	&	-29.0185	&	0.16	\\
35	&	266.4276	&	-28.9982	&	0.19	&	85	&	266.4095	&	-29.0018	&	0.21	&	135	&	266.4246	&	-28.9954	&	0.27	&	185	&	266.4280	&	-29.0100	&	0.25	\\
36	&	266.4150	&	-28.9930	&	0.17	&	86	&	266.4051	&	-28.9855	&	0.33	&	136	&	266.4140	&	-29.0012	&	0.21	&	186	&	266.4259	&	-29.0001	&	0.19	\\
37	&	266.4172	&	-29.0046	&	0.19	&	87	&	266.4219	&	-29.0137	&	0.13	&	137	&	266.3889	&	-29.0212	&	0.22	&	187	&	266.4299	&	-29.0278	&	0.13	\\
38	&	266.4174	&	-28.9965	&	0.24	&	88	&	266.4129	&	-29.0107	&	0.18	&	138	&	266.4082	&	-28.9907	&	0.24	&	188	&	266.3913	&	-28.9923	&	0.23	\\
39	&	266.4174	&	-28.9965	&	0.24	&	89	&	266.4168	&	-28.9858	&	0.28	&	139	&	266.3993	&	-29.0272	&	0.19	&	189	&	266.4055	&	-29.0183	&	0.26	\\
40	&	266.4109	&	-29.0149	&	0.23	&	90	&	266.4199	&	-29.0245	&	0.07	&	140	&	266.4063	&	-29.0262	&	0.09	&	190	&	266.3894	&	-29.0011	&	0.26	\\
41	&	266.4201	&	-29.0093	&	0.14	&	91	&	266.4264	&	-28.9882	&	0.20	&	141	&	266.4080	&	-29.0158	&	0.18	&	191	&	266.3941	&	-29.0131	&	0.16	\\
42	&	266.3907	&	-28.9865	&	0.43	&	92	&	266.4253	&	-28.9931	&	0.27	&	142	&	266.3992	&	-28.9940	&	0.31	&	192	&	266.4369	&	-29.0173	&	0.22	\\
43	&	266.3942	&	-29.0258	&	0.18	&	93	&	266.4036	&	-29.0204	&	0.19	&	143	&	266.4073	&	-29.0024	&	0.23	&	193	&	266.4146	&	-28.9995	&	0.17	\\
44	&	266.4144	&	-29.0073	&	0.19	&	94	&	266.4240	&	-29.0035	&	0.19	&	144	&	266.4356	&	-28.9908	&	0.25	&	194	&	266.3896	&	-28.9938	&	0.38	\\
45	&	266.4267	&	-28.9974	&	0.19	&	95	&	266.3921	&	-28.9873	&	0.37	&	145	&	266.4102	&	-29.0262	&	0.15	&	195	&	266.3903	&	-29.0097	&	0.19	\\
46	&	266.4123	&	-29.0274	&	0.15	&	96	&	266.4171	&	-29.0075	&	0.18	&	146	&	266.4232	&	-28.9971	&	0.20	&	196	&	266.4107	&	-29.0191	&	0.17	\\
47	&	266.4312	&	-29.0014	&	0.16	&	97	&	266.4157	&	-29.0120	&	0.13	&	147	&	266.4298	&	-28.9910	&	0.28	&	197	&	266.4228	&	-29.0179	&	0.10	\\
48	&	266.4027	&	-29.0041	&	0.18	&	98	&	266.3948	&	-29.0216	&	0.25	&	148	&	266.4252	&	-29.0054	&	0.19	&	198	&	266.4076	&	-29.0214	&	0.16	\\
49	&	266.4033	&	-29.0247	&	0.15	&	99	&	266.4191	&	-28.9876	&	0.24	&	149	&	266.4121	&	-29.0136	&	0.15	&	199	&	266.3984	&	-29.0092	&	0.16	\\
50	&	266.4216	&	-29.0107	&	0.15	&	100	&	266.3908	&	-29.0189	&	0.20	&	150	&	266.4351	&	-29.0155	&	0.17	&	200	&	266.4221	&	-29.0198	&	0.13	\\

\hline
\end{tabular}
 \begin{flushleft}
\begin{footnotesize}
The source numbers of the calibration sources and the test group stars: 1: GCIRS 7, 2: GCIRS 1W, 3: GCIRS 3, 4: GCIRS 13, 6: GCIRS 21, 13: GCIRS 29, 14: GCIRS 6E, 21: GCIRS 9, 31: GCIRS 16 NE and 96: GCIRS 16C.
\end{footnotesize}
\end{flushleft}
    \end{center}
  \end{table*}

Optical depths vary over a relatively small range across the field, from 0.07 to 0.43, with a mean value of 0.20$\pm$0.06. We next checked whether systematic biases may have affected the determination of the 3.4 $\,\mu$m optical depth by dividing the sources into four separate categories based on their fluxes, and examining the resulting optical depths. The mean optical depths and standard deviation in each of the 4 quartiles are shown in Table \ref{tab:9}. The differences between them are not significant within the errors. 

\begin{figure*}
  \begin{center}
    \begin{tabular}{cc}
      {\includegraphics[angle=0,scale=0.42]{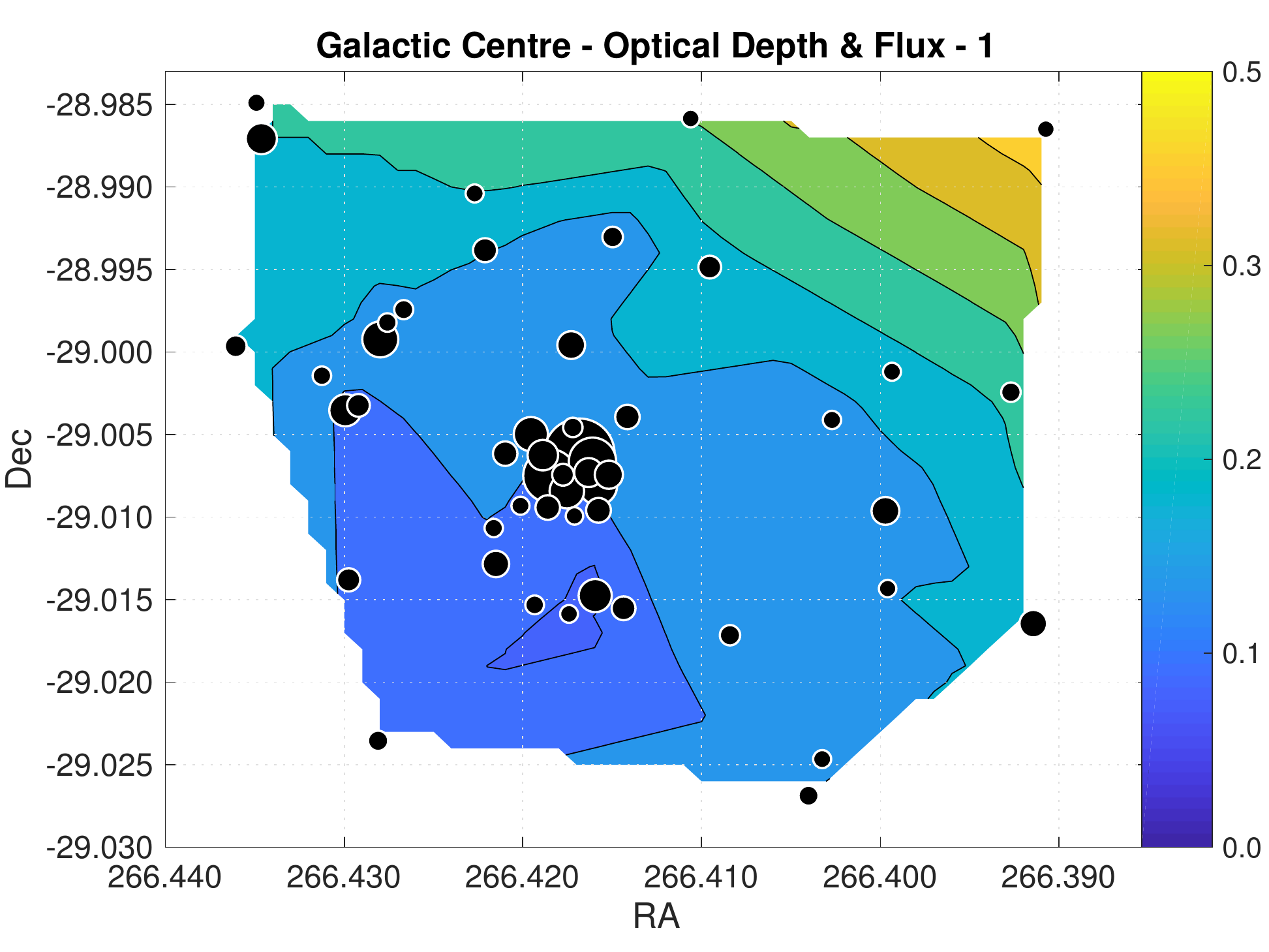}} &
      {\includegraphics[angle=0,scale=0.42]{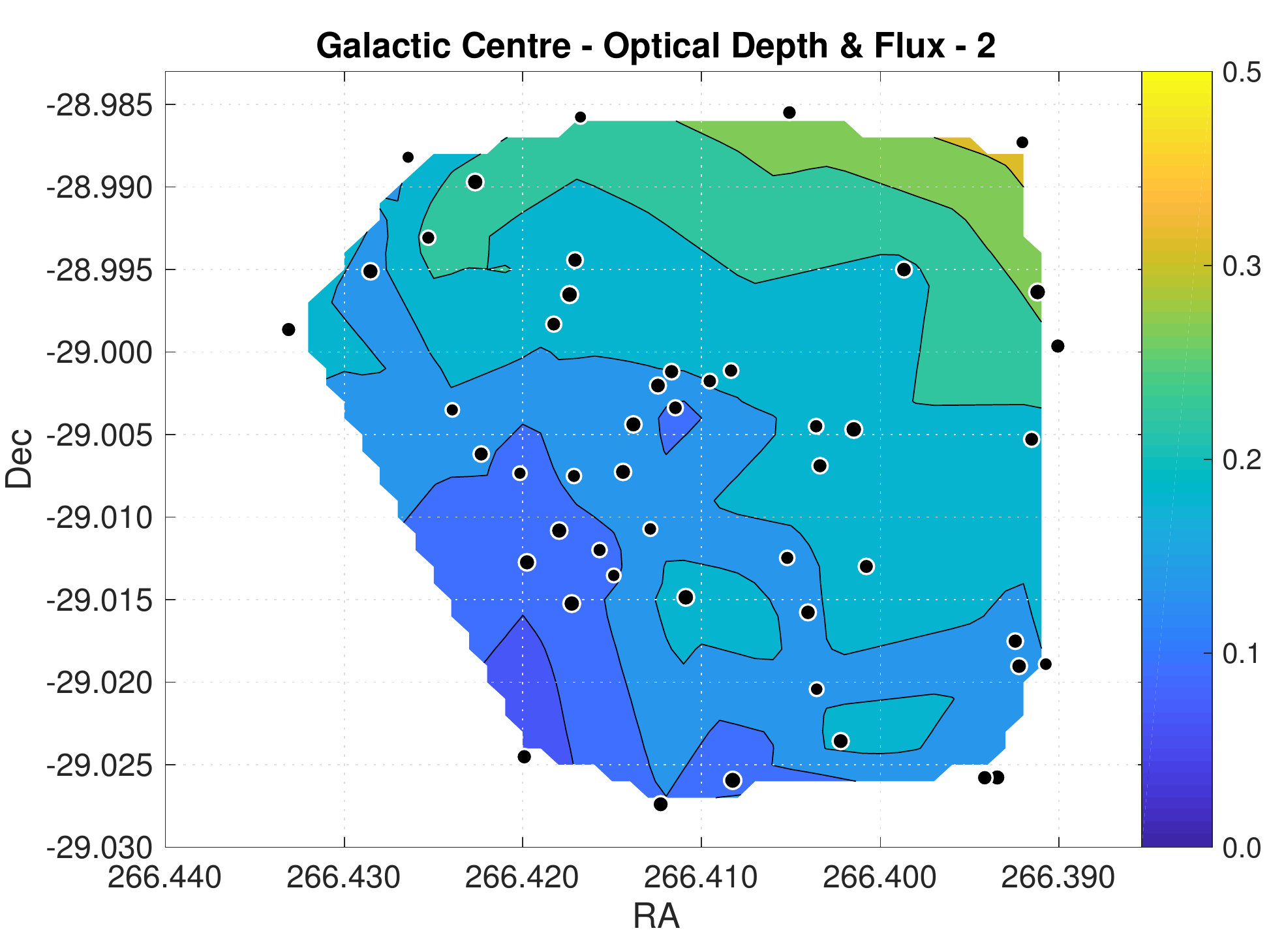}} \\
      {\includegraphics[angle=0,scale=0.42]{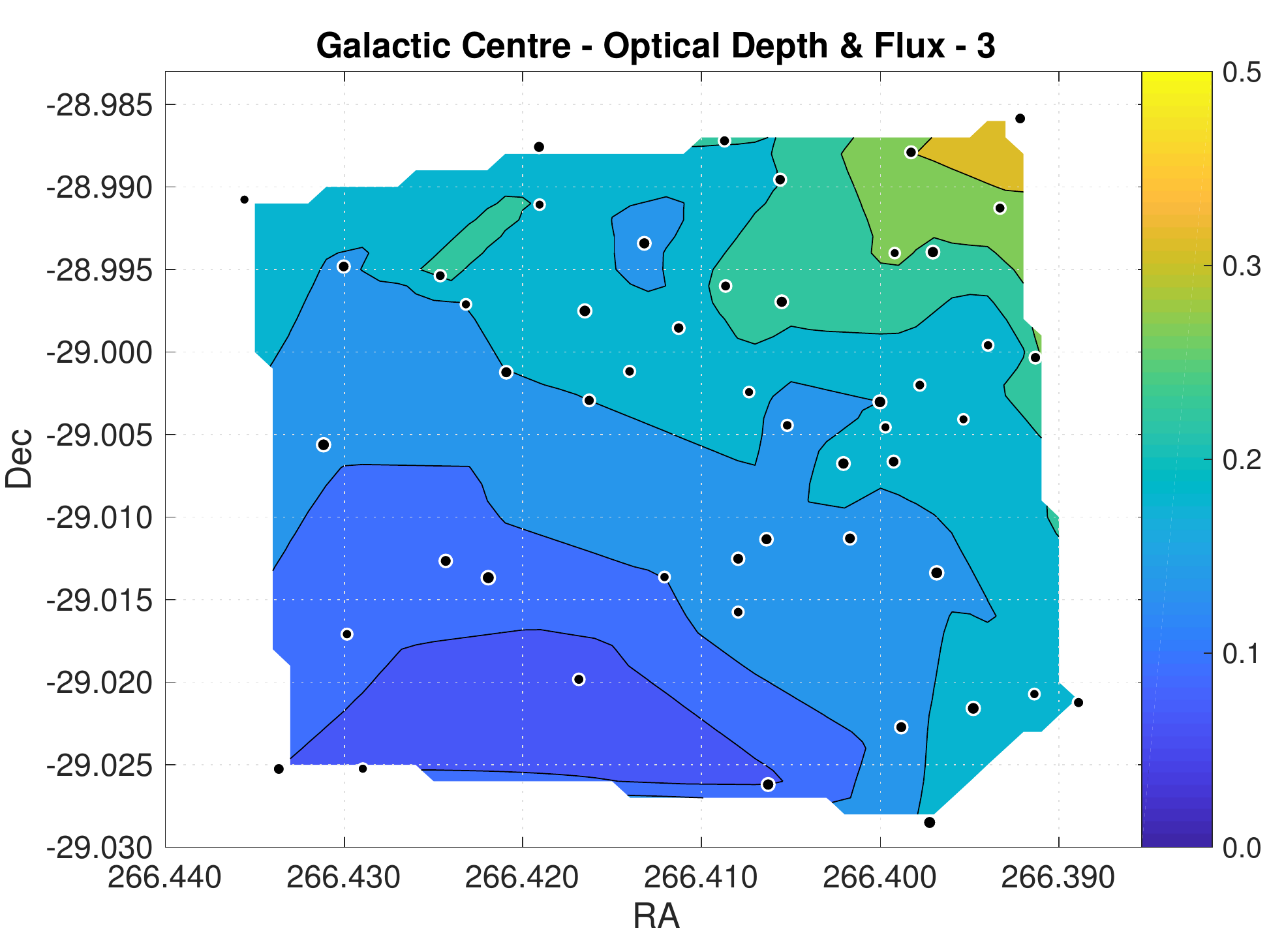}} &
      {\includegraphics[angle=0,scale=0.42]{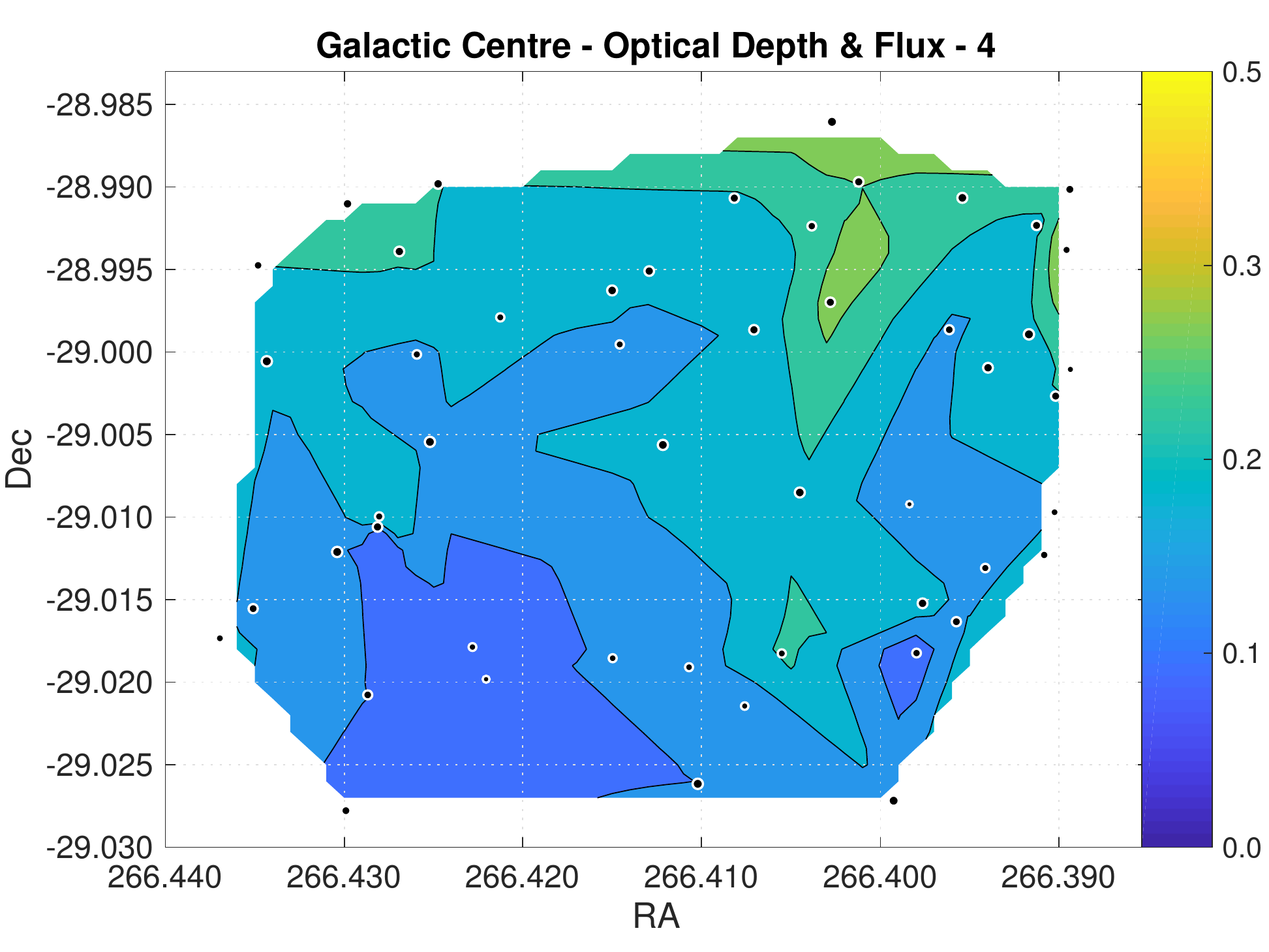}} \\
    \end{tabular}
    \caption{Optical depth maps in four quartiles, each containing 50 sources whose fluxes are selected from the highest to the lowest. The first group contains the brightest sources (Flux - 1) presented in the upper left panel and the last group contains the faintest sources (Flux - 4) presented in the lower right panel. Colour bars and contours indicate the optical depth levels. The location of the sources are shown by black dots whose sizes are proportional to their flux values.}   
   
    \label{fig6}
      \end{center}
\end{figure*}

We then consider the maps of optical depth for the 50 sources in each quartile range based on the flux at 3.6 $\,\mu$m according to order of the sources in Table \ref{tab:7}. These maps are shown in Figure \ref{fig6}. They show optical depth values at 3.4 $\,\mu$m according to the celestial coordinates (J2000). The colour bars and contours represent the optical depth levels. Here the circles represent the stars with their size proportional to their flux. While there is clearly some scatter in the distribution as there are areas within the field with few sources, the same pattern is apparent in each of the quartile ranges. There is a gradient across the field from SE to NW. This is independent of the brightness of individual sources, and thus not likely to arise from the larger uncertainties in optical depth derived from the weaker sources. The overall pattern is similar between the four plots, and there does not appear to be any significant variation between them associated with the brightness of the sources. There is also no significant difference between the clusters and the sources elsewhere in the field of view. At a sensitivity level of $\sim$10$^{-18}$ W\,cm$^{-2}$ $\mu$m$^{-1}$ at 3.6 $\,\mu$m ($\sim$9.5 mag) we conclude that we are able to reliably measure the optical depth using this technique of imaging through narrow band filters.

 \begin{table*}
 \begin{center}
 \footnotesize
  \caption{The mean fluxes ($\times$\,10$^{-18}$ W\,cm$^{-2}$ $\mu$m$^{-1}$) and mean 3.4 $\mu$m optical depths in each of the 4 quartiles.}
  \label{tab:9}
  \centering 
 \begin{tabular}{| c | c | c | c | c | }
     \hline 
    & Flux Group 1 &   Flux Group 2 &   Flux Group 3 &   Flux Group 4 \\
  \hline 
  Flux  & 	16.94	&	5.62	&	3.22	&	1.97	\\	
  \hline 
$\tau_{3.4\,\mu m}$ & 0.18$\pm$0.05  & 0.20 $\pm$0.06 & 0.21 $\pm$0.07  & 0.21$\pm$0.06	\\

\hline
\end{tabular}
\end{center}
\end{table*}

\begin{figure*}
  \begin{center}
    \begin{tabular}{cc}
      {\includegraphics[angle=0,scale=0.42]{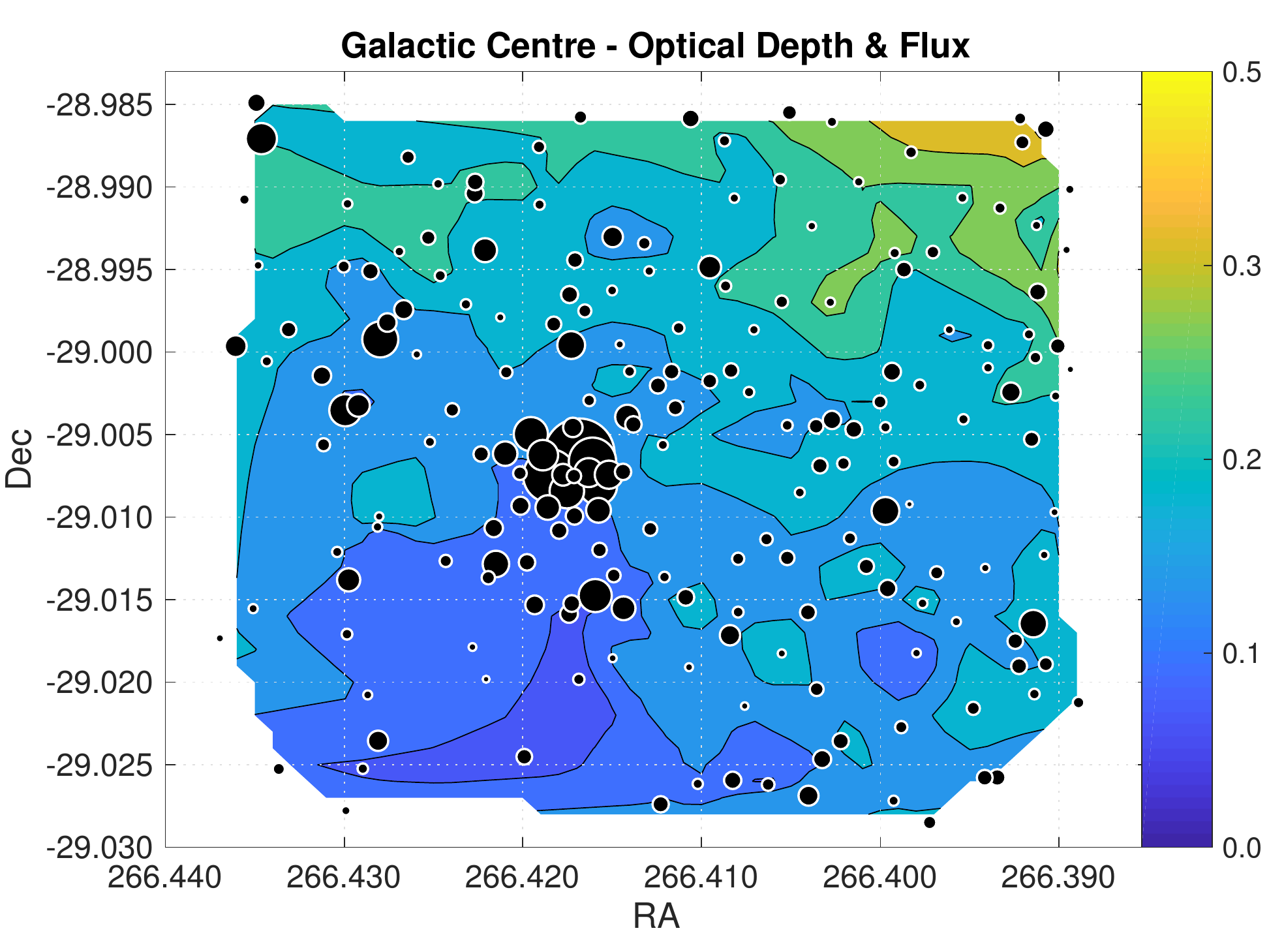}} &
      {\includegraphics[angle=0,scale=0.42]{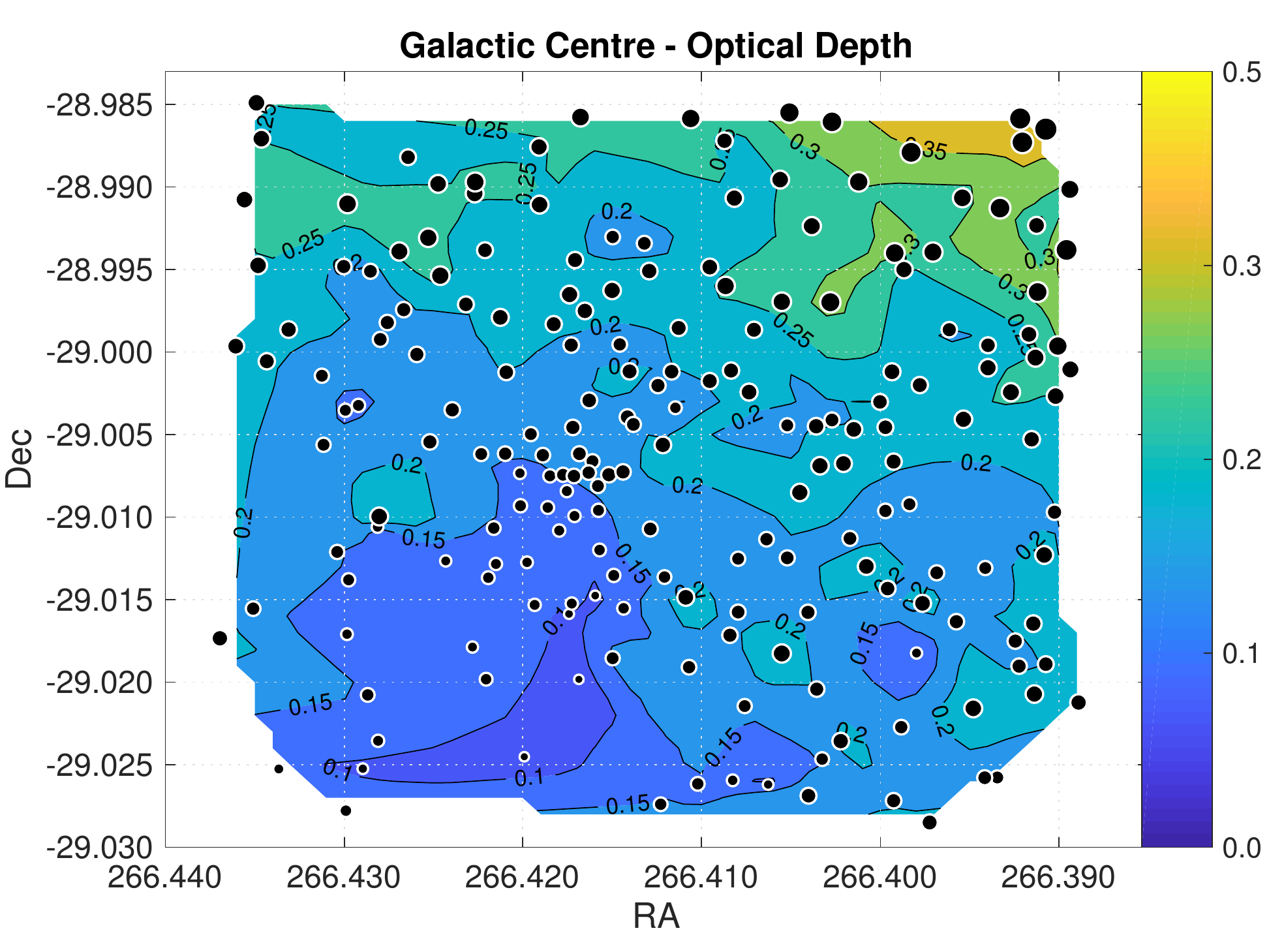}} \\
        \end{tabular}
    \caption{Map of the optical depth from 3.4$\,\mu$m absorption feature for the Galactic Centre in celestial coordinates. Colour bars and contours indicate the optical depth levels. On the left panel location of the sources are shown by black dots whose sizes are proportional to the flux values. On the right panel instead, the sizes are proportional to the 3.4 $\,\mu$m optical depth.}   
    \label{fig7}
      \end{center}
\end{figure*}

We thus take all 200 sources and plot their optical depth in Figure \ref{fig7}. The colours bar and contours show the optical depth at 3.4 $\,\mu$m. On the left panel, the size of the circles is proportional to the flux. On the right panel the size of the circles is proportional to the 3.4 $\mu$m optical depth. 

There could readily be local source-to-source variations in optical depth, superimposed on a broader trend. However, on examination of Figure \ref{fig7} (right panel), this does not appear to be the case. The optical depth, in general, rises from $\sim$0.1 to $\sim$0.4 moving SE to NW across the field at the centre of the Galaxy, with little inter-source scatter.

We then calculated aliphatic hydrocarbon column densities based on the $\tau_{3.4\,\mu m}$ values presented in Table \ref{tab:8} by applying Equation 1 and using the aliphatic hydrocarbon absorption coefficient \textit{A} = $4.7\times10^{-18}$\,cm\,group$^{-1}$ and bandwidth of 62\,cm$^{-1}$ for the 3.4$\,\mu$m filter. Normalised aliphatic hydrocarbon abundances (C/H) (ppm) were calculated based on gas-to-extinction ratio $N(H) = 2.04 \times 10^{21}$\,cm$^{-2}$ mag$^{-1}$ \citep{Zhu2017}, by assuming A$_{V}$$\sim$30 mag.

The column density of aliphatic hydrocarbon rises from $\sim$$8.9\times10^{17}$\,cm$^{-2}$ to $\sim$$5.7\times10^{18}$\,cm$^{-2}$ for $\tau_{3.4\,\mu m}$ ranging from 0.07 to 0.43, with a corresponding abundance with respect to H of $\sim$15\,ppm to $\sim$93\,ppm in the data set. A mean value of $\tau_{3.4\,\mu m}$ $\sim$ 0.2 corresponds to a typical column density of aliphatic hydrocarbon of $2.6\times10^{18}$\,cm$^{-2}$ and 43\,ppm aliphatic hydrocarbon abundance. Comparing this to the ISM total carbon abundance of 358\,ppm \citep{Sofia2001}, there is approximately 12$\%$ of carbon in aliphatic form. 

\begin{figure*}
  \begin{center}
    \begin{tabular}{cc}
      {\includegraphics[angle=0,scale=0.44]{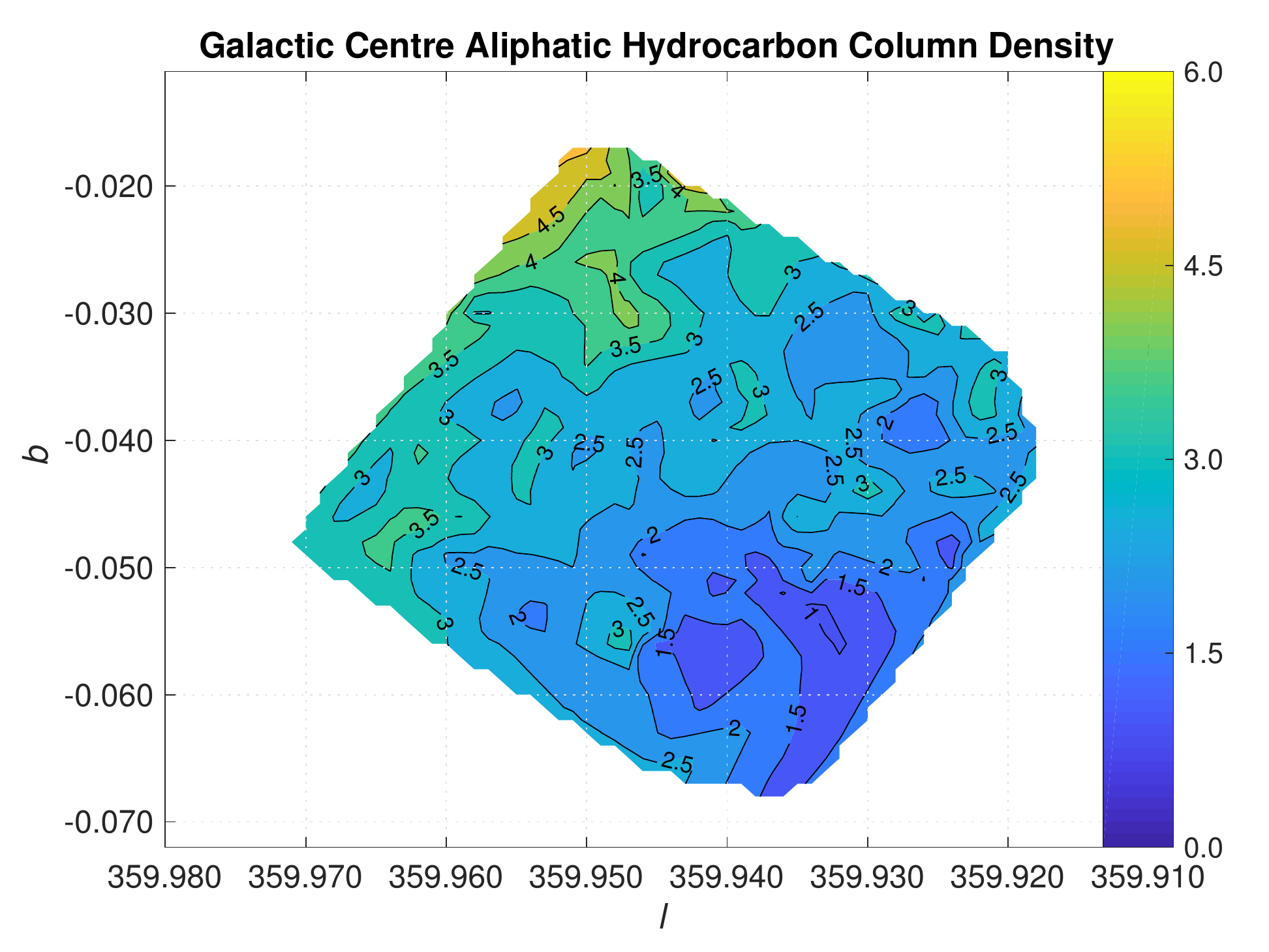}} &
      {\includegraphics[angle=0,scale=0.44]{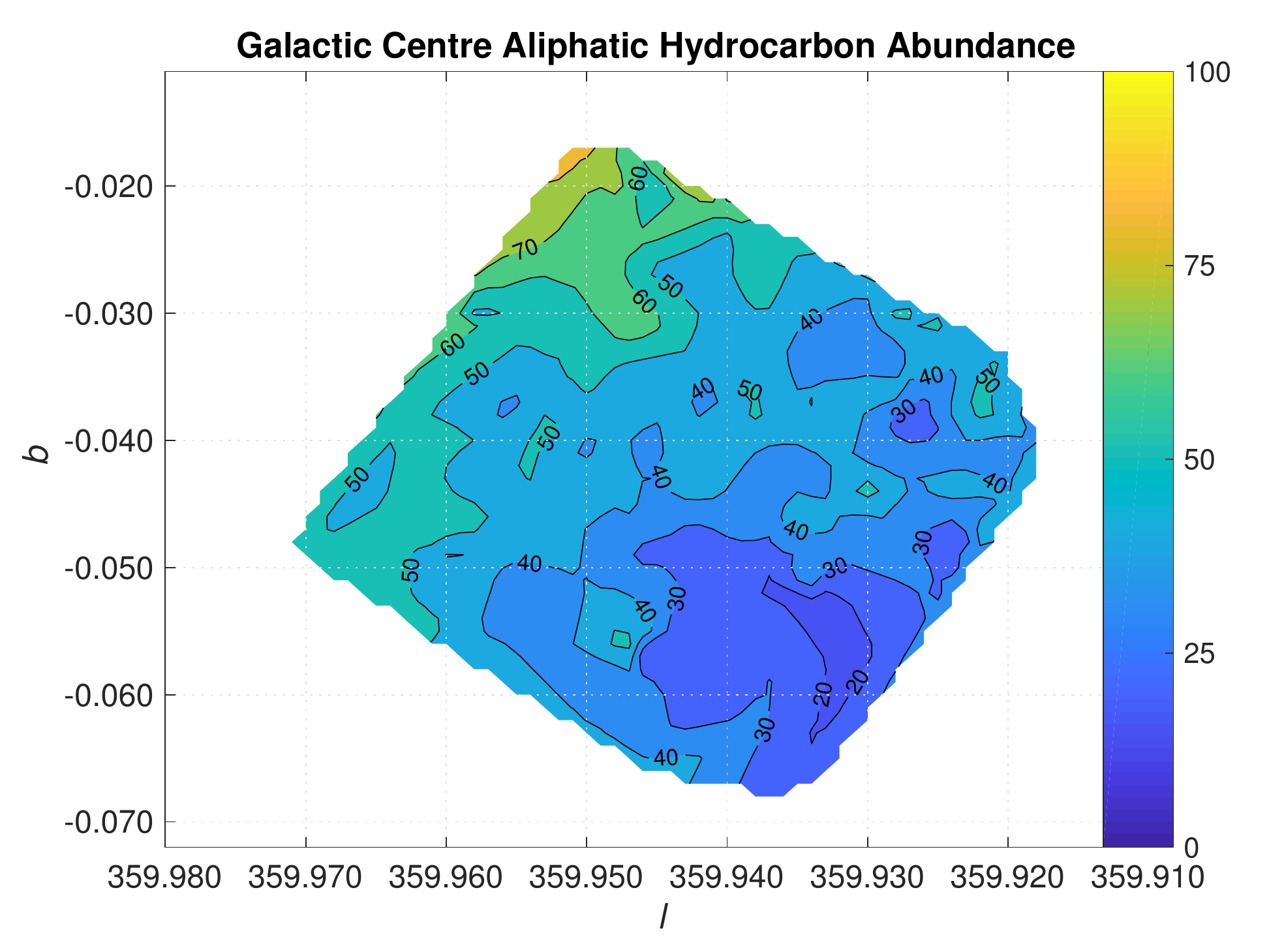}} \\
        \end{tabular}
    \caption{Map of the aliphatic hydrocarbon column density is presented in the left panel for the Galactic Centre in galactic coordinates. Colour bar and contours indicate the  column density ($\times$10$^{18}$ cm$^{-2}$) levels. Map of the aliphatic hydrocarbon abundance relative to H abundance (A$_{V}$$\sim$30 mag) is presented in the right panel for the Galactic Centre in galactic coordinates. Colour bar and contours indicate the aliphatic hydrocarbon abundance levels  (ppm).} 
     \label{fig8}
      \end{center}
\end{figure*}

Finally, to show the column density and abundance gradient in the galactic plane we converted the celestial coordinates presented in Table \ref{tab:8} into galactic coordinates. The resultant aliphatic hydrocarbon column densities and abundances are shown in Figure \ref{fig8}. In the maps, column densities and normalised abundances (which are calculated based on invariant A$_{V}$) appear to increase toward $b$=0, the centre of the Galactic plane.

\section{Discussion and Conclusions} \label{Section6}

The principal result of this work is that the optical depth at 3.4\,$\mu$m can be reliably measured using flux measurements through a series of narrow band filters placed across the feature and the 3.4\,$\mu$m L-band, at least to the flux level of the faintest sources measured in this field in the centre of the Galaxy. This shows that large scale mapping of the aliphatic carbon content can be undertaken by imaging through a series of narrow band filters.

The absorption could occur locally around each source. It could be dominated by absorption along the sight line, or it could arise from a combination of both. If absorption was determined principally by source characteristics it would be expected to vary considerably from source to source. This has not been observed, with the optical depth varying only by a factor of 6 between the very smallest and the largest of all the 200 sources measured. This suggests, therefore, that the optical depth being measured is a general property of the interstellar medium along the sight line to the Galactic centre, rather than being associated with individual sources. The measurements, therefore, trace bulk properties of the interstellar clouds, in particular the amount of aliphatic carbon they contain. 

In the direction of the Galactic centre the optical depth was found to be typically $\sim$ 0.2, corresponding to about $2.6\times10^{18}$\,cm$^{-2}$ and 43$\,$ppm aliphatic hydrocarbon abundance, approximately one tenth of the total carbon. Scaling errors arising from uncertainty in the total column (derived from the value for the visual extinction) dominate the error in the the absolute level of the ppm derived. However the errors in the optical depth at 3.4\,$\mu$m derived for each source, based on the variation in the determinations made using different calibration sets, is likely less than 0.06.

Across the field, aliphatic hydrocarbon column densities are seen to be increasing toward the centre of the Galactic plane. Aliphatic hydrocarbon abundances also appear to be rising in the same direction. However we need to note that abundances were calculated assuming A$_{V}$ is constant so that the gas density is invariant across the field. Therefore further investigations considering A$_{V}$ maps and changes in gas density are required.

We have extended this study to two other fields, one elsewhere in the Galactic centre where the 3.4\,$\mu$m absorption has not previously been measured, the other towards a field in the Galactic plane. We will report on these results in a subsequent paper.

\section*{Acknowledgments}

BG would like to thank to The Scientific and Technological Research Council of Turkey (T\"{U}B\.{I}TAK) as our work has been supported with 2214/A International Research Fellowship Programme. TWS is supported by the Australian Research Council (CE170100026 and DP190103151).

 \bibliographystyle{mn2e}{}
 \bibliography{List}

\end{document}